\documentclass{aa}  
\usepackage{graphicx}
\usepackage{color}
\usepackage{url}
\usepackage[colorlinks=true,citecolor=blue,linkcolor=blue]{hyperref}
\usepackage[normalem]{ulem}
\usepackage{CJK}
\usepackage{amsmath}
\usepackage{fancyhdr}
\usepackage{threeparttable}
\renewcommand{\markboth}[2]{}
\renewcommand{\markright}[1]{}
\renewcommand{\markboth}[2]{}
\renewcommand{\markright}[1]{}
\usepackage{txfonts}
\titlerunning{Multi-wavelength study of EP250416a / GRB 250416C}
\usepackage{natbib}
\newcommand{\gtap}{\mathrel{\hbox{\rlap{\lower.55ex \hbox {$\sim$}}
                   \kern-.3em \raise.4ex \hbox{$>$}}}}
\newcommand{\ltap}{\mathrel{\hbox{\rlap{\lower.55ex \hbox {$\sim$}}
                   \kern-.3em \raise.4ex \hbox{$<$}}}}

\def\be{\begin{equation}} 
\def\ee{\end{equation}}

\begin{document}
\begin{CJK*}{UTF8}{gbsn}

\title{Multi-wavelength study of EP250416a / GRB 250416C: An Optically Dark Long GRB with a Late Jet Break}

\author{Guoying Zhao (赵国英)\inst{1,2} $^\dagger$
        \and
        Duo-Le Cao (曹多乐)\inst{1,2} $^\dagger$
        \and
        Rong-Feng Shen (申荣锋)\inst{1,2}
        \thanks{Corresponding author \\ $^\dagger$ These authors contributed equally to this work.}
       \and
      Hui Sun\inst{3}
             \and
   Chi-Chuan Jin\inst{3}
    \and
    Wei-Min Yuan\inst{3}
    \and
    Chen-Wei Wang\inst{4}
    \and
    Shao-Lin Xiong\inst{4}
    \and
    Dmitry Svinkin\inst{5}
    \and
    Dong Xu \inst{3,6}
    \and
    Shuai-Qing Jiang\inst{3,6}
    \and
    Peter G. Jonker\inst{7}
    \and
    Yun Wang\inst{8,9}
    \and
    Hao Zhou\inst{8}
    \and
    Chang Zhou\inst{10}
    \and
    Xinlei Chen\inst{11}
    \and
    Kaushik Chatterjee\inst{11}
    \and
    Xue-Feng Wu\inst{8}
    \and
    Xiao-Feng Wang\inst{12}
    \and
    Chun Chen\inst{1,2}
    \and
    Yuan Liu\inst{3}
    \and
    Andrew J. Levan\inst{7,13}
    \and
    Jennifer Alexandra Chacon Chavez\inst{14}
    \and
    Jonathan Quirola-V\'asquez\inst{7}
    \and
    Franz E. Bauer\inst{15}
    \and
    Antonio Martin-Carrillo \inst{16}
    \and
    Gregory Corcoran\inst{16}
    \and
    Daniele B. Malesani\inst{7,17,18}
    \and
    Dmitry Frederiks\inst{5}
    \and
    Anna Ridnaia\inst{5}
    \and
    Alexandra L. Lysenko\inst{5}
    \and
    Mikhail Ulanov\inst{5}
    \and 
    Anastasia Tsvetkova\inst{5}
          }

   \institute{School of Physics and Astronomy, Sun Yat-Sen University, Zhuhai, 519082, P. R. China \and CSST Science Center for the Guangdong-Hongkong-Macau Greater Bay Area, Sun Yat-Sen University, Zhuhai, 519082, P. R. China
      \email{zhaogy28@mail2.sysu.edu.cn, shenrf3@mail.sysu.edu.cn}
               \and
National Astronomical Observatories, Chinese Academy of Sciences, Beijing 100101, China
           \and
State Key Laboratory of Particle Astrophysics, Institute of High Energy Physics, Chinese Academy of Sciences, Beijing, 100049, China
              \and
Ioffe Institute, Politekhnicheskaya 26, 194021 St. Petersburg, Russia
\and
School of Astronomy and Space Science, University of Chinese Academy of Sciences, Chinese Academy of Sciences, Beijing 100049, People's Republic of China
\and
Department of Astrophysics/IMAPP, Radboud University, P.O. Box 9010, 6500 GL, Nijmegen, The Netherlands
\and
Purple Mountain Observatory, Chinese Academy of Sciences, Nanjing 210023, China
\and
Department of Astronomy, University of California, Berkeley, CA 94720-3411, USA
\and
Department of Astronomy, School of Physics, Huazhong University of Science and Technology
\and
South-Western Institute for Astronomy Research, Yunnan University
\and
Physics Department, Tsinghua University, Beijing, 100084, China
\and
Department of Physics, University of Warwick
\and
Instituto de Astrof{\'{\i}}sica, Facultad de F{\'{i}}sica, Pontificia Universidad Cat{\'{o}}lica de Chile, Campus San Joaquín, Av. Vicuña Mackenna 4860, Macul Santiago
\and
Instituto de Alta Investigaci{\'{o}}n, Universidad de Tarapac{\'{a}}, Casilla 7D, Arica, Chile
\and
School of Physics and Centre for Space Research, University College Dublin, Belfield, Dublin 4, Ireland
\and
The Cosmic Dawn Centre (DAWN), Denmark
\and
Niels Bohr Institute, University of Copenhagen, Jagtvej 155, DK-2200, Copenhagen N, Denmark
}

\date{Received xx  / Accepted xx}           

 \abstract{We present multi-wavelength study of the $\gamma$/X-ray transient EP250416a (also designated GRB 250416C), triggered by the Einstein Probe (EP) Wide-field X-ray Telescope and also by SVOM and Konus-Wind (KW). Observations spanning the $\gamma$-ray, X-ray, and optical bands facilitated detailed analysis of the burst's prompt emission, afterglow evolution, and physical origin. GRB 250416C exhibits a burst duration of 30 s in X-ray and 17.7 s in $\gamma$-rays, with joint spectral fitting of 0.5-5000 keV data gives $E\rm_{peak}=342_{-232}^{+90}$ keV. Optical spectroscopy of the afterglow, acquired with the Gemini  Multi-Object Spectrograph (GMOS) on Gemini South, yielded a redshift of $z=0.963$.  Accounting for the measured redshift, the isotropic energies are $E\rm_{X,iso}=2.7_{-0.5}^{+0.9}\times10^{50}$ erg and $E\rm_{\gamma,iso}=7.34_{-2.1}^{+5.1}\times10^{51}$ erg, aligning with the Amati relation for long GRBs. The fluence ratio $\rm S(25-50~keV)/S(50-100~keV)=0.78_{-0.12}^{+0.10}$ classifies GRB 250416C as an X-ray rich (XRR) GRB. The X-ray afterglow shows an initial shallow decay ($\alpha \approx -0.5$) transitioning to a canonical decay phase ($\alpha \approx -1$), with a very late jet break at $t\sim 1.5\times 10^6$ s, corresponding to a jet half-opening angle of $\theta _j=10.6_{-1.8}^{+1.9}$ degrees. GRB 250416C is optically dark, as it shows only a faint $r$-band detection ($r=24.16\pm 0.13$ mag) from Gemini South-GMOS and a low optical-to-X-ray spectral index $\beta_{\rm OX} = 0.3$. This may be attributed to significant host-galaxy extinction, with a required $A_V^{\text{host}}=2.7\pm 0.3\ \text{mag}$ derived from the extinction curve model.}

\maketitle
%%%%%%%%%%%%%%%%%%%%%%%%%%%%%%%
\section{Introduction} 

Gamma-ray bursts (GRBs) are the most powerful explosions since the Big Bang~\citep{1973ApJ...182L..85K}. They exhibit two stages: an initial prompt $\gamma$-ray emission phase and a long-lasting multi-wavelength afterglow emission phase~\citep{2004IJMPA..19.2385Z, 2007ChJAA...7....1Z,2015PhR...561....1K}. These sudden, random events manifest as intense flashes of $\gamma$-ray radiation originating from cosmological distances, often accompanied by high X-ray fluxes and typically lasting $\leq$1000 seconds \citep{1993ApJ...413L.101K}. A key observational characteristic of GRBs is their duration in the $\gamma$-ray band, quantified by the parameter T$_{90}$, defined as the time interval encompassing $5 \sim 95 \%$ of the cumulative $\gamma$-ray flux. Based on T$_{90}$, GRBs are empirically classified into two distinct populations: long GRBs ( $\gtrsim$ 2 s) and short GRBs ( $\lesssim$ 2 s) ~\citep{1993ApJ...413L.101K}. The physical origins of these two partially overlapping classes have been firmly established through decades of multi-wavelength and multi-messenger observations. Long GRBs originate from the collapse of massive stars, often accompanied by supernovae, while short GRBs arise from compact binary mergers ~\citep{2009ApJ...703.1696Z}. 

The light curves of GRB prompt emissions exhibit pronounced variability and irregularity, reflecting the temporal characteristics of internal energy dissipation and the activity of the central engine \citep{1994ApJ...430L..93R}. In addition, the spectral properties of GRBs encode crucial information about the underlying radiation mechanisms \citep{1996ApJ...459..393N}. Spectrally, prompt emission is typically modeled as a cutoff power law (CPL), with a photon index $\alpha$ describing the low-energy slope and a peak energy E$_\mathrm{peak}$ marking the transition to a steeper high-energy cutoff. The Amati relation, a tight and well-established empirical correlation between the rest-frame E$_\mathrm{peak}$ and the isotropic-equivalent prompt energy E$_\mathrm{iso}$, unifies long GRBs and provides a key diagnostic for measuring their cosmological distance and scaling their energies~\citep{2002A&A...390...81A}. Additional clues to prompt  emission physics come from temporal lags between light curves in different energy bands, which encode information about the emission region's size, expansion speed, and radiation mechanism \citep{2006ChJAA...6..312Z,2017ApJ...834L..13W}.

Following the prompt phase, the afterglow emission dominates GRB detectability, persisting from hours to weeks (or even months) across X-ray, optical, and radio bands. Afterglows are generated by the deceleration of relativistic ejecta as they interact with the surrounding circumburst medium- either a uniform interstellar medium or a wind-like medium shed by the progenitor star~\citep{1992MNRAS.258P..41R, 1993ApJ...405..278M}. This interaction drives a forward shock back into the ejecta; the majority of afterglow emission arises from synchrotron radiation of electron accelerated in the forward shock. A key feature of afterglow light curves is their power-law decay, while jet breaks, critical transitions in the light curve, occur when the relativistic beaming angle of the jet exceeds the physical opening angle, causing a sudden steepening of the decay index~\citep{1999ApJ...519L..17S,2002ApJ...568..820G}.These jet breaks are invaluable for constraining the jet's physical opening angle, performing de-beaming correction on the isotropic-equivalent, and accurately measuring the true energy output of GRBs.

On 2025 April 16 at 17:53:59 UT, the Wide-field X-ray Telescope (WXT) on board EP detected a bright X-ray flare (0.5- 4 keV), designated EP250416a~\citep{2025GCN.40154....1Z}. The Follow-up X-ray Telescope (FXT) on board EP began observations 130 s after the WXT trigger and detected the source ~\citep{2025GCN.40165....1Z}. Concurrently, KW $\gamma$-ray spectrometer and the Space-based multi-band astronomical Variable Object Monitor (SVOM) Gamma-Ray Monitor (GRM) detected an associated $\gamma$-ray burst GRB 250416C~\citep{2025GCN.40167....1S, 2025GCN.40184....1S}. Subsequent Follow-up by Swift/X-ray Telescope (XRT) and ground-based optical facilities (MASTER-OAFA, GSP, GMOS, and TRT) provided critical X-ray and optical data to constrain the burst's prompt and afterglow properties~\citep{2025GCN.40166....1S,2025GCN.40157....1L,2025GCN.40156....1L,2025GCN.40219....1L,2025GCN.40160....1L}.

In this paper, we present a comprehensive multi-wavelength analysis and physical interpretation of the afterglow of EP250416a/GRB 250416C, leveraging data from EP, KW, SVOM-GRM, Swift-XRT and ground-based optical telescopes. Throughout the paper, the reference time $\rm T_{\rm 0}$ is defined as $\text{2025-04-16T17:53:29 }$ (UT). The paper is structured as follows. The observations and data reduction are described in Section~\ref{sec:Observation and data reduction}. The light curve and spectral properties of the prompt X-ray and $\gamma-$ray emission are analyzed in Section~\ref{sec:Prompt_emissions}. The multi-band afterglow light curves are investigated in Section~\ref{sec:afterglow_emissions}. Afterglow modeling with energy injection is presented in Section~\ref{sec:modeling}. The physical origin of GRB 250416C, its classification as an optically dark GRB, and other key parameters are discussed in Section~\ref{sec:discussion}. Finally, our conclusions are summarized in Section~\ref{sec:conclusion}. Throughout the paper, a cosmology model of $\Omega _{\rm_M}=0.268$, $\Omega \rm_\Lambda=0.714$, $\rm H_{\rm 0}=69.6 ~\rm km~s^{-1}~Mpc^{-1}$ is used~\citep{2014ApJ...794..135B}.

%%%%%%%%%%%%%%%%%%%%%%%%%%%
\section{Observation and data reduction}\label{sec:Observation and data reduction}

\subsection{Einstein Probe Observations}

The Einstein Probe (EP), successfully launched on 9 January 2024, is a dedicated soft X-ray sky monitoring mission designed to detect and characterize transient phenomena such as GRBs, X-ray novae, and tidal disruption events~\citep{2022hxga.book...86Y}. The mission is equipped with two complementary scientific instruments. Operating in the soft X-ray band (0.5-4.0 keV), WXT is optimized for discovering fast transients, boasting an exceptionally wide instantaneous field of view that enables continuous, unbiased coverage of a large fraction of the X-ray sky. FXT is optimized for targeted, high-resolution follow-up observations of detected transients, operating in the 0.3-10 keV band. It comprises two identical modules (FXT-A/B), each equipped with 54 nested Wolter-I mirrors and a PN-CCD detector. The detector supports three readout modes (Full Frame, Partial Window, Timing) to balance temporal resolution and sensitivity.

\begin{figure}[ht!]
\centering        
\includegraphics[width=9.0cm]{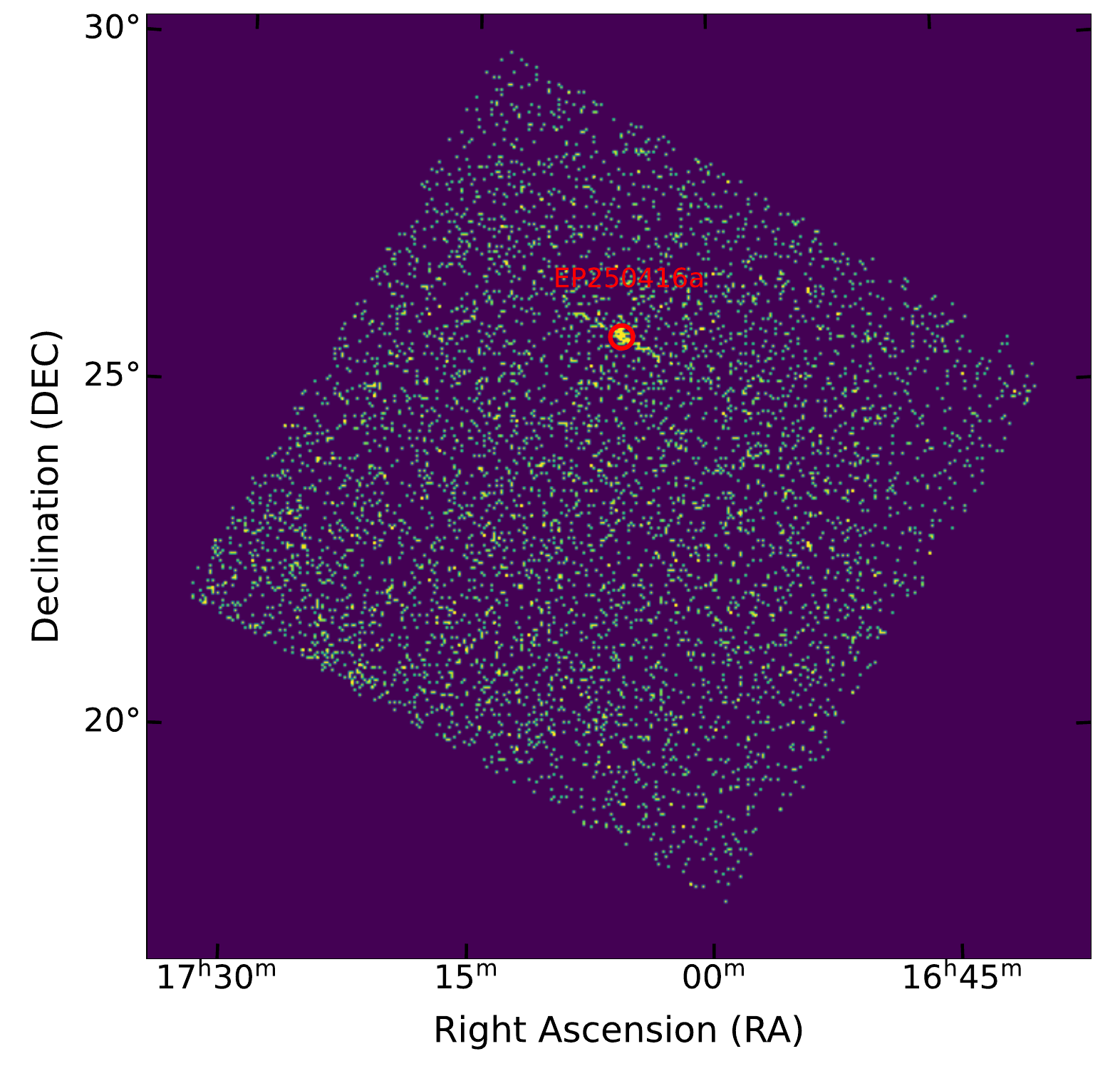}
\caption{Image of EP250416a captured by EP-WXT CMOS17 detector at 17:53:59 UTC on 16 April 2025. This frame corresponds to the initial EP-WXT trigger detection of the event. The source is localized at R.A. = 256.42$^\circ$ and Dec. = 25.78$^\circ$, providing the initial sky position used for subsequent follow-up analyses.}
\label{fig:X-ray_image}
\end{figure}

EP250416a was detected by the WXT at 17:53:59 UTC on 16 April 2025, with an initial trigger position of R.A. = 256.42$^\circ$ and Dec. = 25.78$^\circ$ (J2000) ~\citep[see Figure~\ref{fig:X-ray_image}]{2025GCN.40154....1Z}. This position was later verified by FXT and Swift-XRT observations, ensuring high positional accuracy for multi-wavelength follow-up. 

Following the WXT trigger, FXT initiated its first observation of EP250416a at 130 s after the burst, yielding a refined localization at R.A. $= 256.4228^{\circ}$ and Dec.$= 25.7755^{\circ}$. A total of seven observational epochs were carried out, spanning from 130 s to approximately 23 days after the burst. This temporal coverage successfully captures both the early shallow-decay phase and the subsequent normal decay phase of the X-ray afterglow.

The WXT data were processed using \texttt{wxtpipeline}, the standard reduction pipeline of the WXT Data Analysis Software, together with the corresponding calibration database (CALDB)~\citep{2025ExA....60...15C}, while the FXT data were reduced using the \texttt{fxtchain} tool within the FXT Data Analysis Software (\texttt{FXTDAS})\footnote{\url{https://epfxt.ihep.ac.cn/analysis}}. Spectral analysis was performed with \texttt{Xspec}\footnote{\url{https://heasarc.gsfc.nasa.gov/xanadu/xspec/manual/manual.html}} for EP data.
 
%%%%%%%%%%%%%%%%%%%%%%%%%%%%

%%%%%%%%%%%%%%%%%%%%%%%%%%%%
\subsection{Swift-XRT Observations}
Swift/XRT is a key facility for GRB afterglow follow-up, offering high sensitivity in the 0.3-10 keV band and rapid response capabilities~\citep{2004HEAD....8.1603B}. Swift-XRT performed two targeted observations of GRB 250416C: the first at $\sim$ 4.8 hours (PI: Kennea) and the second at $\sim$ 20.5 hours (PI: EPSC) after the WXT trigger. These observations complemented the EP/FXT data by covering a time window when the afterglow was transitioning from the shallow-decay to the canonical decay phase.

Data reduction was performed using the XRT Data Analysis Software (\texttt{XRTDAS v3.6.1}) with the standard xrtpipeline. From the Level 2 event files, source and background spectra were extracted with \texttt{xselect}, adopting a circular source region of $20''$ centered on GRB 250416C. Ancillary response files (ARFs) were produced with xrtmkarf, while the corresponding response matrix files (RMFs) were obtained from the CALDB database. Finally, All spectra were grouped using \texttt{GRPPHA} for subsequent analysis. 

\subsection{Konus-Wind and SVOM-GRM Observations}

The Konus-Wind instrument (KW) is a $\gamma$-ray spectrometer consisting of two identical detectors, S1 and S2, which observe the southern and northern ecliptic hemispheres, respectively~\citep{1995SSRv...71..265A}. It detected GRB 250416C in waiting mode at 17:53:47 UTC on 16 April 2025, at $\mathrm{T_{\rm 0}}+18$ s~\citep{2025GCN.40167....1S}. In waiting mode, KW records count rates with a coarse time resolution of 2.944 s resolution across three energy bands: G1(18-75 keV), G2 (75-311 keV), and G3 (311-1231) keV. Bayesian blocks decomposition of the 18-311 keV (G1+G2) light curve where the burst is most prominent reveals a $\sim 7\sigma$ count-rate increase in the interval from $\rm T_{\rm 0}+8.809$ s to $\rm T_{\rm 0}+26.473$ s.

The Gamma-Ray Monitor (GRM) is the high-energy detector aboard the Chinese-French satellite the Space-based multi-band astronomical Variable Object
Monitor (SVOM) which is dedicated to $\gamma$-Ray Burst studies~\citep{2009AIPC.1133...25G,2010arXiv1005.5008S}. SVOM-GRM detected GRB 250416C as a weak single-pulse burst~\citep{2025GCN.40184....1S}, with temporal characteristics consistent with KW data.

\subsection{Gemini South-GMOS} \label{sec:gmos}

\begin{figure}[ht!]
\centering
    \includegraphics[width=9cm]{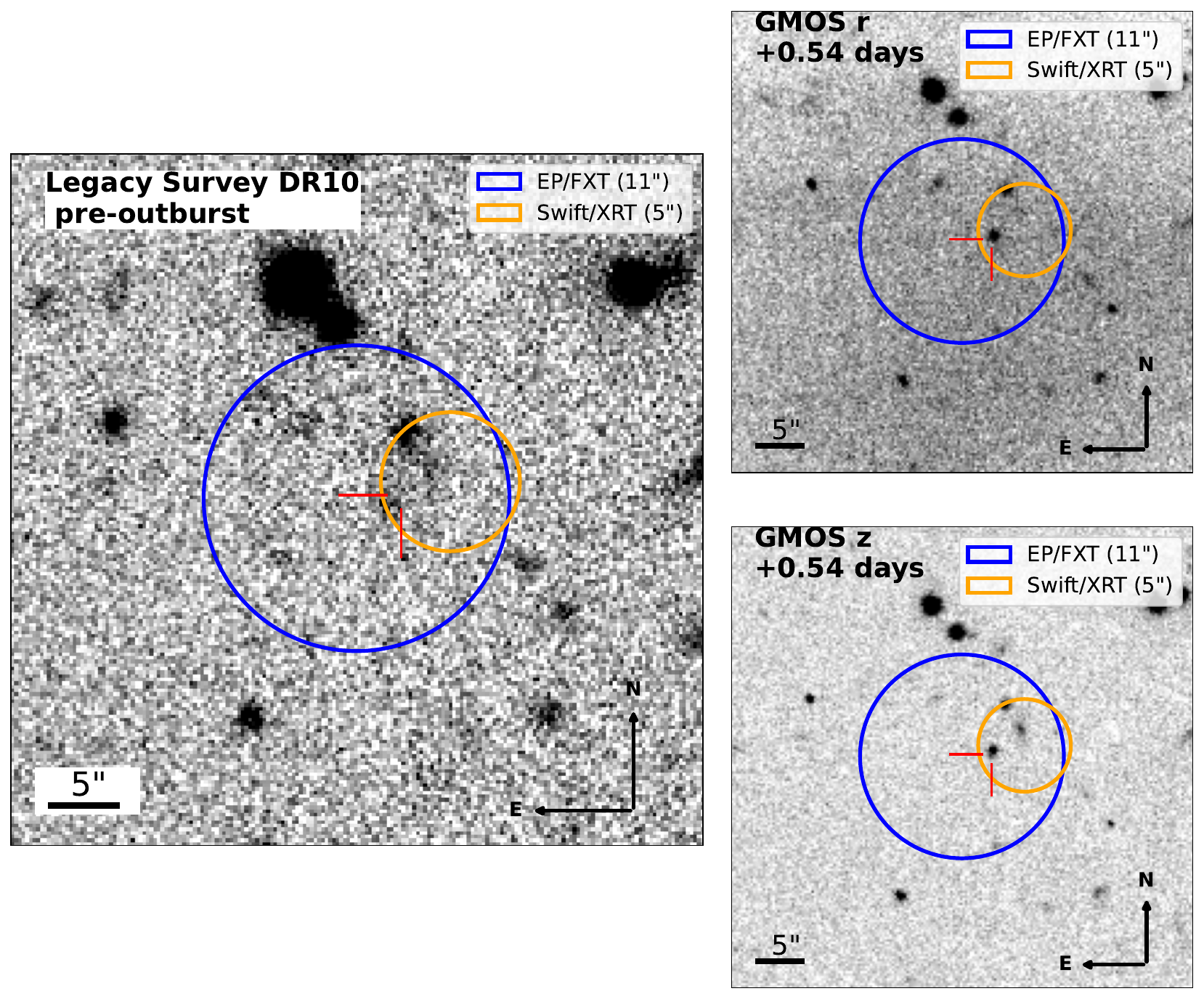}
    \caption{Pre-outburst LS DR10 image near the counterpart of GRB 250416C (left panel), and post-outburst images of the counterpart detected with Gemini-GMOS in the $r$-band (upper right) and $z$-band (lower right) at 0.54 days after the outburst. Orange and blue lines denote the positional uncertainties of Swift/XRT and EP/FXT, respectively, and the red mark indicates the detection position of the counterpart by GMOS in the target field.}
    \label{fig:LS+GMOS_image}
\end{figure}

Gemini South Observatory's GMOS instrument conducted follow-up observations of the counterpart of GRB 250416C. Imaging observations were performed $\sim 0.54$ days post-trigger with the $r$ filter (four 60-s exposures), detecting an optical source within the EP/FXT and Swift/XRT localization region with an AB r-band magnitude of $22.8\pm 0.1$ (combining GRB afterglow and host galaxy emission) ~\citep{2025GCN.40160....1L}. The position of the source, calibrated with WCS, is $\rm R.A. = 17:05:40.83$, $\rm Dec. = +25:46:31.5$, consistent with the X-ray and EP/FXT positions (see Figure~\ref{fig:LS+GMOS_image}). 

Notably, pre-outburst observations of the region near the GMOS detection position were obtained by the Legacy Survey DR10 (LS DR10), and the image shows a faint galaxy at this position (i.e., the subsequently confirmed host galaxy,see Figure~\ref{fig:LS+GMOS_image}) with an $r$-band magnitude of $23.16\pm 0.11$\footnote{Note that an initial uncertainty estimate of 0.17 mag was obtained from a nearby SDSS galaxy identified in the VizieR catalog. We subsequently derived an uncertainty of 0.11 mag from the $r$-band flux and the inverse variance provided by the LS DR10 Tractor catalog, which we adopt here.} and a photometric redshift of $0.72 \pm 0.24$ (\citealt{2019AJ....157..168D,2021MNRAS.501.3309Z}). In contrast, the optical source detected by GMOS at the same position after the outburst has a magnitude of $22.8\pm 0.1$, which is brighter than the faint pre-outburst galaxy. This confirms that the optical source detected after the outburst is the optical counterpart and host galaxy of GRB 250416C, and there is a brightness enhancement phenomenon after the outburst. 

To isolate the optical afterglow emission, we converted the GMOS-measured magnitude ($r = 22.8\pm 0.1$) and the LS DR10 host-only magnitude ($23.16\pm 0.11$) to fluxes and did the subtraction. This yields an afterglow magnitude of $r=24.16\pm 0.13$ for GRB 250416C, confirming its optically faint nature.

\begin{figure}[ht!]
    \centering
    \includegraphics[width=9cm]{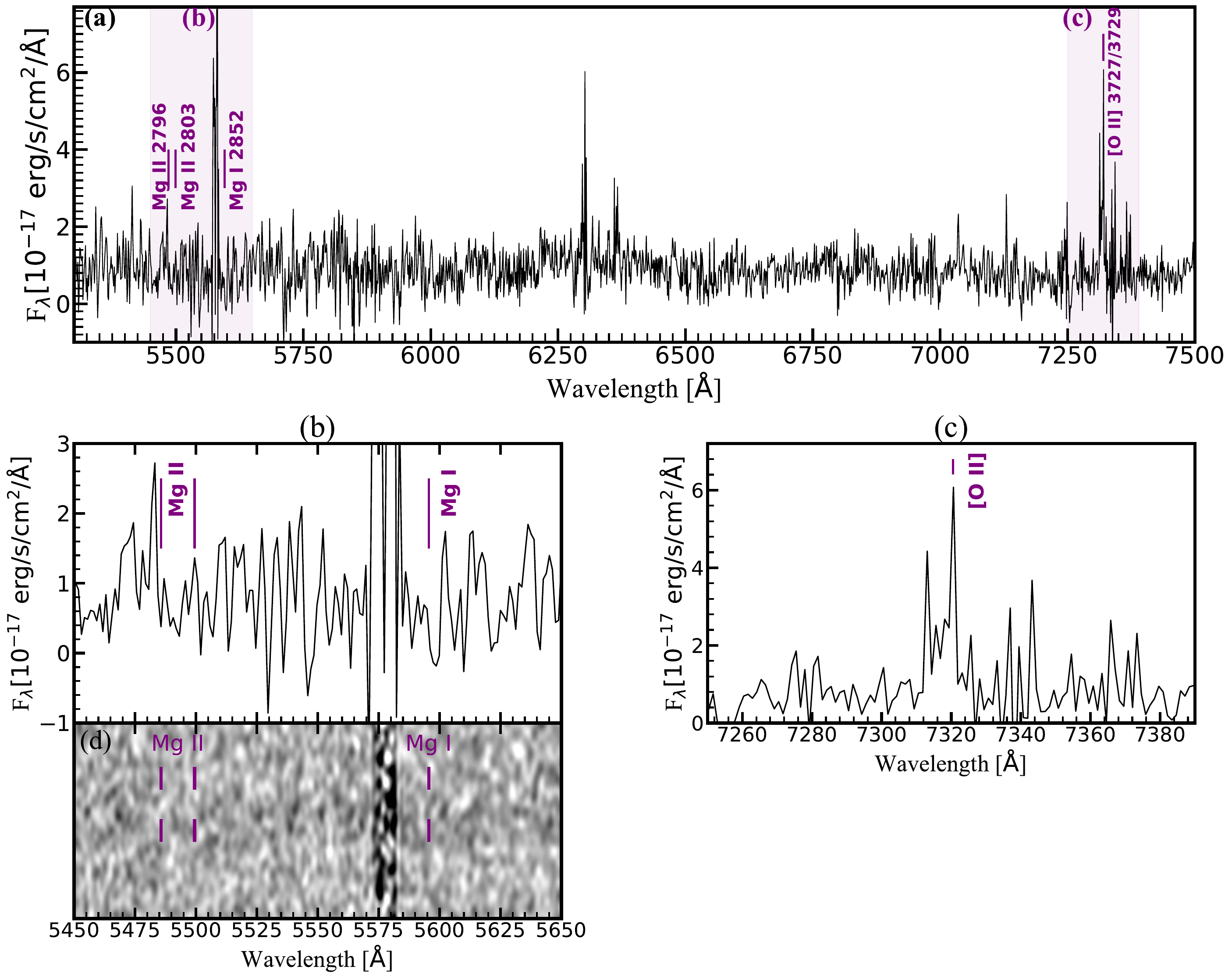}
    \caption{
    From the observed emission lines, particularly the [O II] doublet $\lambda 3727, 3729$, we derive a redshift of $z=0.963$ for GRB 250416C, assuming the transient event occurred within its host galaxy at this redshift. The lower two panels provide zoomed views: panel (b) covers the spectral range around the tentative Mg II ($\lambda_{\rm rest}=2796,2803$~\AA) and Mg I ($\lambda_{\rm rest}=2852$~\AA) features at $z\approx0.962$, while panel (c) focuses on the [O II] emission peak. Panel (d) shows the corresponding two-dimensional spectrum.} %\textcolor{red}{(RFS: please check the wavelength locations of the vertical-line marks for Mg I in panels (b) and (d) )}}
    \label{fig:GMOS_spec}
\end{figure}

Spectroscopic observations were carried out $~0.58$ days post-trigger using the B480 grating (four 900-s exposures, wavelength range $\sim 4000-9500 \text{\AA}$, ~\cite{2025GCN.40219....1L}). A prominent emission line at $7317 \text{\AA}$ is identified as the unresolved $[\rm O II] \lambda3727/\lambda3729$ doublet, a canonical host galaxy feature (Figure~\ref{fig:GMOS_spec}). Using the relation \(z=(\rm \lambda_{\rm obs}-\lambda_{\rm rest})/\lambda_{\rm rest}\), we derive a redshift of \(z=0.963\) for the host galaxy, corresponding to a luminosity distance of 6.4 Gpc. 

Tentative absorption features consistent with Mg II ($\lambda 2796$, $\lambda 2803$) and Mg I ($\lambda 2852$) at \(z=0.962\) are present but not definitive, due to low signal-to-noise ratios. To further assess these features, we inspect the smoothed two-dimensional spectral trace (Figure~\ref{fig:GMOS_spec} d), which shows a flux gap at the expected position of the blended Mg II doublet. The apparent emission-like peaks near the nominal Mg II wavelengths are also present in the 2D trace, but their separation is inconsistent with the intrinsic doublet spacing and is therefore likely due to noise. Overall, the 2D spectrum provides additional support for a weak Mg II/Mg I absorption system at $z=0.962$.

\vspace{-0.3cm}

\subsection{Further Optical Photometric Observations}

We obtained images with the telescope of the Thai Robotic Telescope network (TRT) located at Cerro Tololo Inter-American Observatory, Chile. Observations obtained at a median exposure time of $\sim$ 13 hours after the EP/WXT trigger yielded upper limits of $g>21.1,\,r>21.0,\,$ and $i>19.9$. %added by sqjiang

Additionally, we observed the target about 34.24 hours after the EP/WXT trigger using the Alhambra Faint Object Spectrograph and Camera (ALFOSC) mounted on the 2.56~m Nordic Optical Telescope (NOT). The location shows a faint source with the magnitude $r=23.42 \pm 0.22$, which is consistent with the magnitude of the known source in the Legacy Survey DR10. 
%added by sqjiang

We complement this with the remaining optical photometric data provided in the General Coordinates Network (GCN) Circulars\footnote{\url{https://gcn.nasa.gov/circulars}}. The 1-meter telescope at the Las Cumbres Observatory node performed i-band observations approximately 8 hours after the EP/WXT trigger, yielding no detection of a new optical source in the co-added images within the EP/FXT error box, with a limiting magnitude of $\sim$ 20 mag ~\citep{2025GCN.40157....1L}. The MASTER-OAFA robotic telescope obtained a series of unfiltered ($C$-band) observations starting at $\sim 10.3$ hr after the EP/WXT trigger, yielding a final coadded upper limit of $r> 18.2$ mag~\citep{2025GCN.40156....1L}. All optical photometric measurements are summarized in Table~\ref{apptable:Optical Photometric Observations}.

%Optical follow-up of EP250416a was critical for constraining the burst's afterglow luminosity, host galaxy properties, and potential classification as an ``optically dark'' GRB. Observations were conducted using multiple ground-based facilities, leveraging the precise localization provided by EP/FXT to target the burst's position.

%%%%%%%%%%%%%%%%%%%%%%%%%%%%%
\section{Prompt emissions }
\label{sec:Prompt_emissions}
\subsection{Light Curves}

\begin{figure}[ht!]
\centering
\includegraphics[width=9cm]{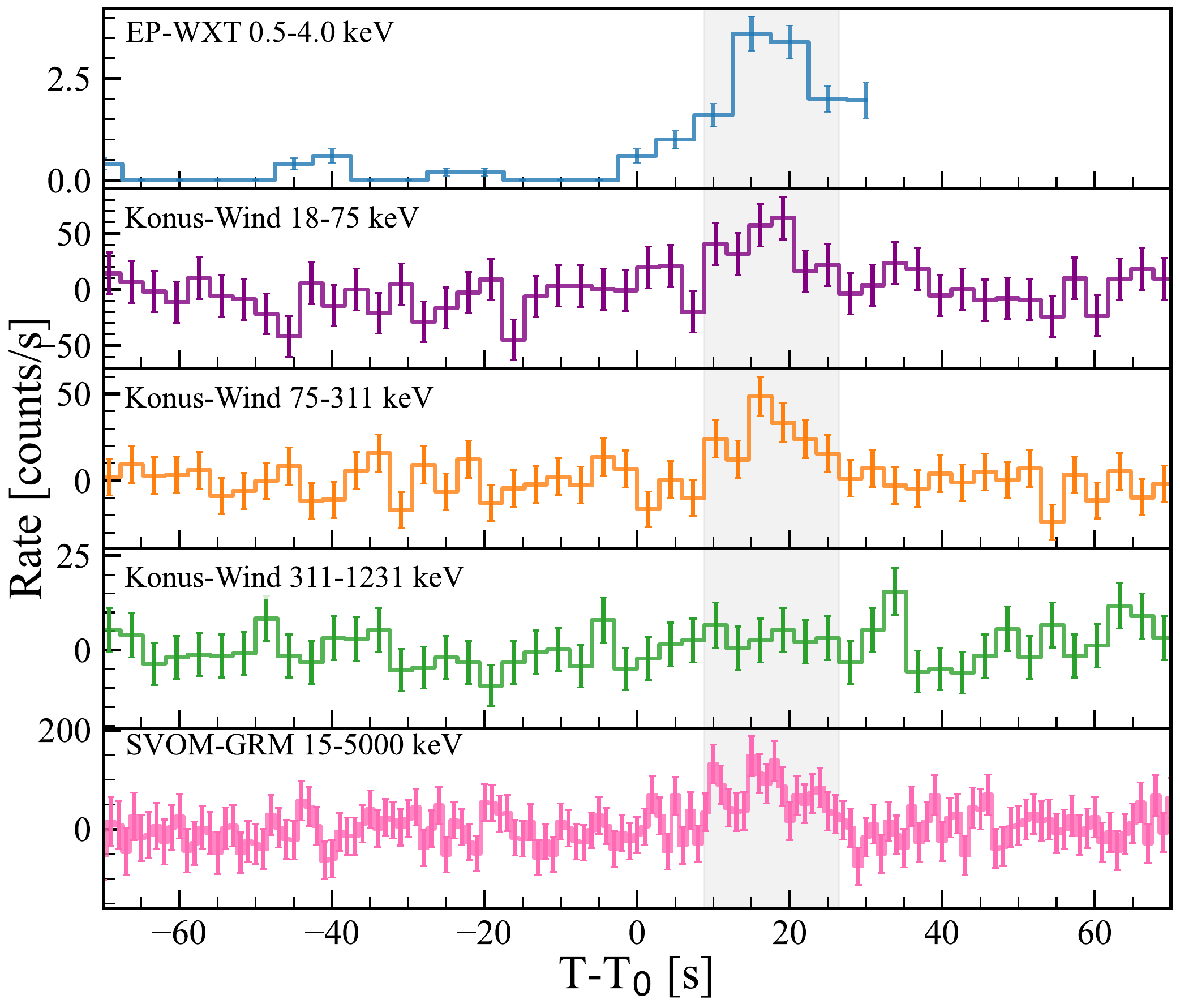}
\caption{High-energy light curves of GRB 250416C. Top panel: EP/WXT light curve in 0.5-4.0 keV band, showing a weak precursor at $\rm T_{\rm 0}-50$ s and a main pulse spanning $\rm T_{\rm 0}$ s to $\rm T_{\rm 0}+30$ s. Middle three panels: KW light curves in 18-75 keV, 75-311 keV and 311-1231 keV bands, with the main $\gamma$-ray pulse peaking at $\rm T_0+18$ s. Lower panel: SVOM-GRM light curve in 15-5000 keV. The shaded gray region marks the $\gamma$-ray emission duration ($\rm T_0+8.809$ s to $\rm T_0+26.473$), and $\rm T_{90,\gamma} \sim$ 17.7 s in the 18-75 keV and 75-311 keV bands. Here, the reference time $\rm T_0$ is defined as $\text{2025-04-16T17:53:29}$. }
\label{fig:WXT+K-W+GRM_lc}
\end{figure}

In  Figure ~\ref{fig:WXT+K-W+GRM_lc}, we present the light curves of GRB 250416C, across X-ray and $\gamma$-ray bands, with consistent temporal features confirming the burst's authenticity and constraining its emission duration.

%%%%%%%%%%%%%%%%%%%%%%%%%%%
%%%%%%%%%%%%%%%%%%%%%%%%

The WXT light curve (top panel of Figure ~\ref{fig:WXT+K-W+GRM_lc}) exhibits two distinct components. First, a weak X-ray precursor detected at $\rm T_0-50$ s, with a duration of $\sim 10$ s and peak flux of $\sim 5\times10^{-10}~\mathrm{erg/s/cm^{2}}$. This is followed by a more intense main pulse spanning $\rm T_0$ s to $\rm T_0+30$ s (total duration $\sim 30$ s) and a peak flux of approximately $10^{-8}~\mathrm{erg/s/cm^{2}}$. 

KW light curves (middle three panels of Figure ~\ref{fig:WXT+K-W+GRM_lc}) in 18-75 keV, 75-311 keV and 311-1231 keV bands show a single, intense $\gamma$-ray pulse peaking at $\rm T_0+18$ s. The $\gamma$-ray emission duration (the shaded region in the Figure ~\ref{fig:WXT+K-W+GRM_lc}) spans $\rm T_0+8.809$ s to $\rm T_0+26.473$ s, with  $\rm T_{90,\gamma}$ is measured as 17.7 s in both the 18-75 keV and 75-311 keV bands. This duration firmly classifies GRB 250416C as a long GRB.

The SVOM-GRM light curve (lower panel of Figure ~\ref{fig:WXT+K-W+GRM_lc}) in the 15-5000 keV band confirms a weak single-pulse profile, with temporal characteristics consistent with KW data. This cross-instrument consistency validates the reliability of the prompt emission's temporal constraints, ruling out instrumental artifacts and confirming the burst's $\gamma$-ray signal across a broad energy range.

\subsection{Joint spectral analysis}

Spectral properties of prompt emission generally reveal the energetic scale of GRBs and constrain the likely radiation mechanism. We performed joint spectral analysis of contemporaneous data from EP/WXT, KW, and SVOM-GRM over the $\gamma$-ray emission interval $\rm T_0+8.809$ s to $\rm T_0+26.473$s, spanning 0.5-5000 keV (Table~\ref{tab:joint_spec_par}). 

\begin{table*}
\caption{The fitting results and corresponding fitting statistics for the prompt emission spectra. The averaged absorbed flux is derived in the 0.5-4.0 keV range for WXT, in the 15-5000 keV range for GRM, in the 18-1231 keV range for KW, and in the 0.5-5000 keV range for the joint fitting of WXT, GRM, and KW. All errors represent the 1$\sigma$ uncertainties.}
\label{tab:joint_spec_par}
    \centering
    \begin{tabular}{*7{c}}
    \hline
    \hline
         Instruments&Time Intervals& Model& Photon Index& E$\rm_{peak}$& Flux& Cstat/dof  \\
         &[Second]& & &[keV]& $\rm erg~cm^{-2}~s^{-1}$ \\
    \hline
    \hline
    WXT& 0-30& PL& $0.35_{-0.15}^{+1.07}$& -&$(7.7\pm2.8)\times 10^{-9}$&16.44/23\\
    \hline
    GRM& 8.809-26.473& CPL& $1.24_{-0.37}^{+0.38}$& $454.7_{-210}^{+160}$& $(8.1\pm 1.3)\times 10^{-8}$& 28.17/16\\
    \hline
    KW& 8.809-26.473& CPL& $1.35_{-0.09}^{+0.09}$& $370_{-138}^{+204}$& $(8.9\pm 1.9)\times 10^{-8}$& 2.63/3\\
    \hline
    WXT+GRM+KW& 8.809-26.473& CPL& $1.43_{-0.05}^{+0.05}$& $342_{-232}^{+90}$& $(7.8\pm 1.5)\times 10^{-8}$& 31.66/45\\
    \hline
    \hline
    \end{tabular}
\end{table*}

The spectra were modeled with an absorbed CPL model (\texttt{tbabs$\times$cutoffpl} in \texttt{Xspec}), which describes the photon flux spectrum as:
\begin{equation*}
    N(E) = A \left(\frac{E}{100~\mathrm{keV} }\right)^{-\alpha} \mathrm{exp}\left[-\frac{E(2-\alpha)}{E_\mathrm{{\rm peak}}}\right],
\end{equation*}
where $A$ is the normalization constant [$\rm phontons~cm^{-2}~s^{-1}~keV^{-1}$], $\alpha$ is the photon index, $E_\mathrm{{peak}}$ is the peak energy of the $E^2N(E)$ spectrum. This model is widely used for GRB prompt emission, as it accounts for both the power-law slope at low energies and the exponential cutoff at high energies. 

\begin{figure}[ht!]
    \centering
    \includegraphics[width=9cm]{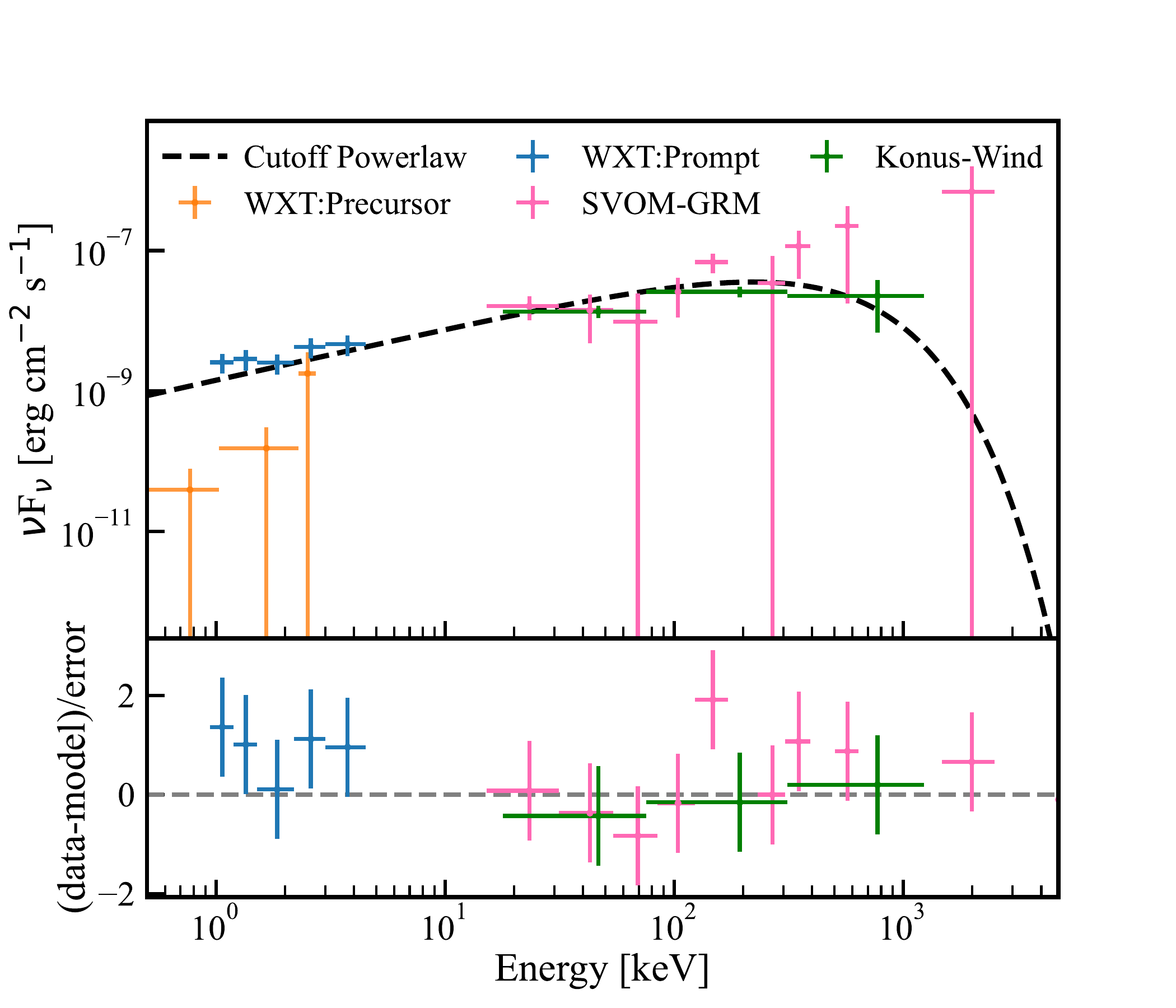}
    \caption{The joint spectral fitting of WXT, Konus-Wind and GRM spectra between $\rm T_0+8.809$ s to $\rm T_0+26.473$ s with an absorbed cutoff power law. The black dash line denotes the best-fit CPL model. For completeness, we also plot the spectrum of the WXT precursor, which is not included in the fit.}   
    \label{fig:WXT+GRM+K-W_spec}
\end{figure}

The joint fit gives \textbf{$\alpha=-1.4\pm 0.1$} and $E_\mathrm{{peak}}=342_{-232}^{+90}\rm keV$. Figure ~\ref{fig:WXT+GRM+K-W_spec} illustrates the spectral fitting results. The upper panel is multi-wavelength data points are overplotted with the best-fit absorbed CPL model. EP/WXT data (0.5-4.0 keV) are shown in orange (precursor; photon index $\Gamma=0.2_{-2.2}^{+1.8}$) and blue (prompt), KW data (18-1231 keV) in green, and SVOM-GRM data (15-5000 keV) in pink. The bottom panel displays the fit residuals, confirming that the CPL model adequately reproduces the multi-wavelength spectral data across all observed energy bands.

%%%%%%%%%%%%%%%%%%%%%%%%%%%%%%%%%

\begin{figure}[ht!]
\centering
\includegraphics[width=9cm]{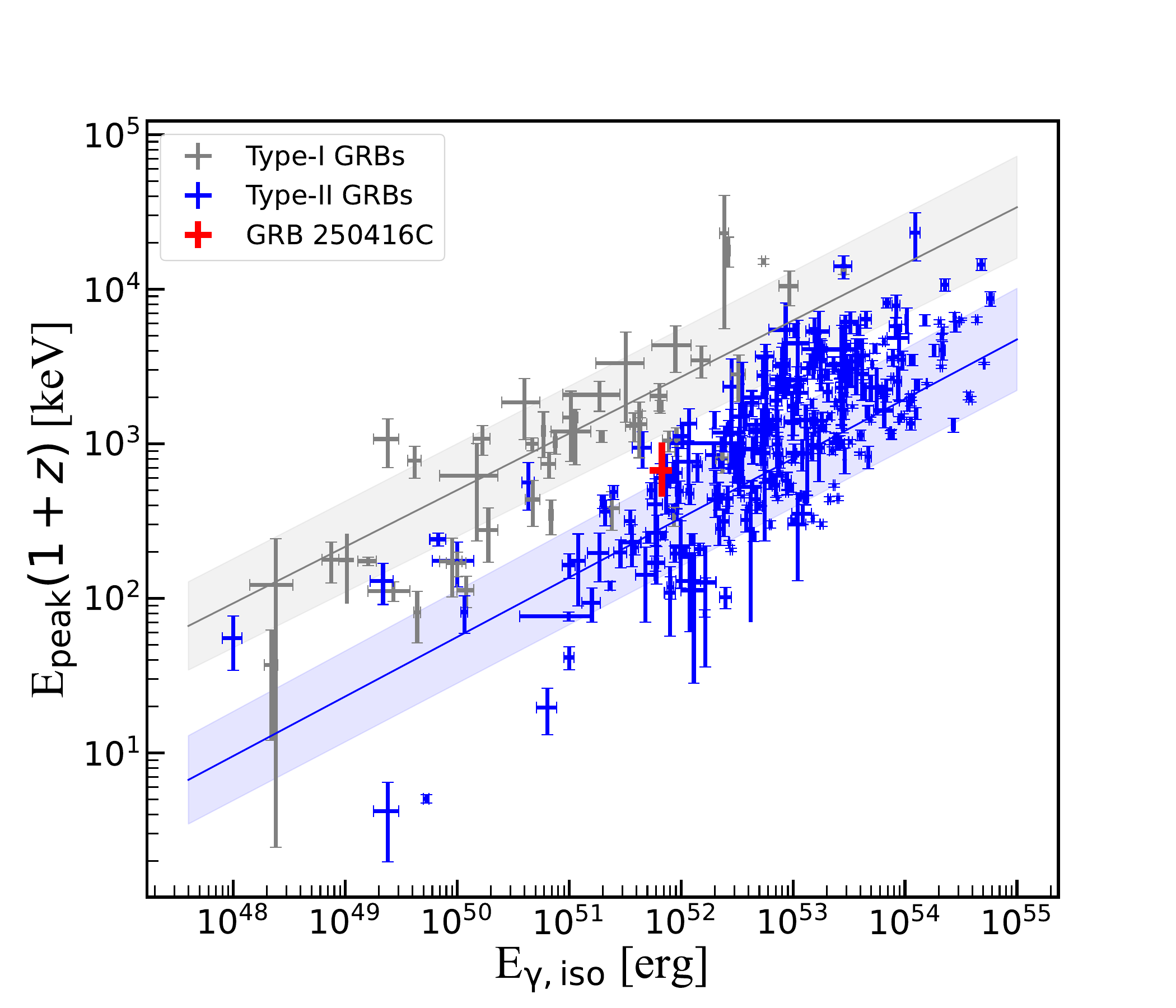}
\caption{The rest-frame spectral peak energies (E$_\mathrm{peak}$) versus the isotropic-equivalent energies in the rest-frame 1--$10^4$ keV band (E$_{\gamma, \mathrm{iso}}$) for GRBs. The best-fit E$_\mathrm{peak}$-E$_{\gamma,\mathrm{iso}}$ (Amati) relation~\citep{2002A&A...390...81A} for both Type $\mathrm{\uppercase\expandafter{\romannumeral1}}$ and Type $\mathrm{\uppercase\expandafter{\romannumeral2}}$ samples are plotted (solid lines) with the 1$\sigma$ scattering bands \citep[shaded regions;][]{2020MNRAS.492.1919M}. The position of GRB 250416C is also marked in the diagram (red cross).}
%\textcolor{red}{(RFS: to be consistent, instead of introducing new symbols, I suggest replacing $E_p$ and $E_{iso}$ in the axis labels by the ones already defined and used in the main text)}}
\label{fig:Epz-Eiso}
\end{figure}

Using the host galaxy redshift z = 0.963 (see Section \ref{sec:gmos}) and corresponding luminosity distance $d_L= 6.4$ Gpc, we calculated the isotropic-equivalent total radiative energies of the prompt emission. From the EP-WXT 0.5-4 keV X-ray data only, we derive $E_{\rm X,iso}= 2.7_{-0.5}^{+0.9}\times 10^{50}$ erg, while from the joint spectral fit of EP-WXT, KW, and SVOM-GRM over 0.5-5000 keV we obtain $E_{\rm \gamma, iso}= 7.34_{-2.1}^{+5.1} \times 10^{51}$ erg. For consistency with conventions, we extrapolated the best-fit spectral model to the rest-frame 1--$10^4$ keV band and obtained $E_{\gamma,{\rm iso}}=8.12_{-2.4}^{+5.6} \times 10^{51}$ erg, allowing GRB 250416C (marked as a red cross) to be plotted on the Amati relation ~\citep{2002A&A...390...81A} diagram (Figure~\ref{fig:Epz-Eiso}), which compares the rest-frame spectral peak energy and the isotropic-equivalent energy. Its position aligns with the distribution of typical (type~\uppercase\expandafter{\romannumeral 2}) long GRBs, further validating its classification and confirming consistency with the empirical spectral-energy relation for long GRBs. 

%%%%%%%%%%%%%%%%%%%%
%%%%%%%%%%%%%%%%%%%%%

We also classified GRB 250416C using the fluence ratio criterion proposed by \cite{2008ApJS..175..179S} for distinguishing classical GRBs (C-GRBs), X-ray rich GRBs (XRRs) and X-ray flashes (XRFs). The criterion is defined by the ratio of fluences in the 25-50 keV and 50-100 keV bands:
\begin{equation}
\begin{aligned}
        &S(25-50\ \text{keV})/S(50-100\ \text{keV})\leq 0.72, &(\text{C-GRB}),\\ 
    0.72 < &S(25-50\ \text{keV})/S(50-100\ \text{keV})\leq1.32,&(\text{XRR}),\\
    1.32< &S(25-50\ \text{keV})/S(50-100\ \text{keV}), &( \text{XRF}).
\end{aligned}
\end{equation}

\begin{figure}[ht!]
    \centering
    \includegraphics[width=9cm]{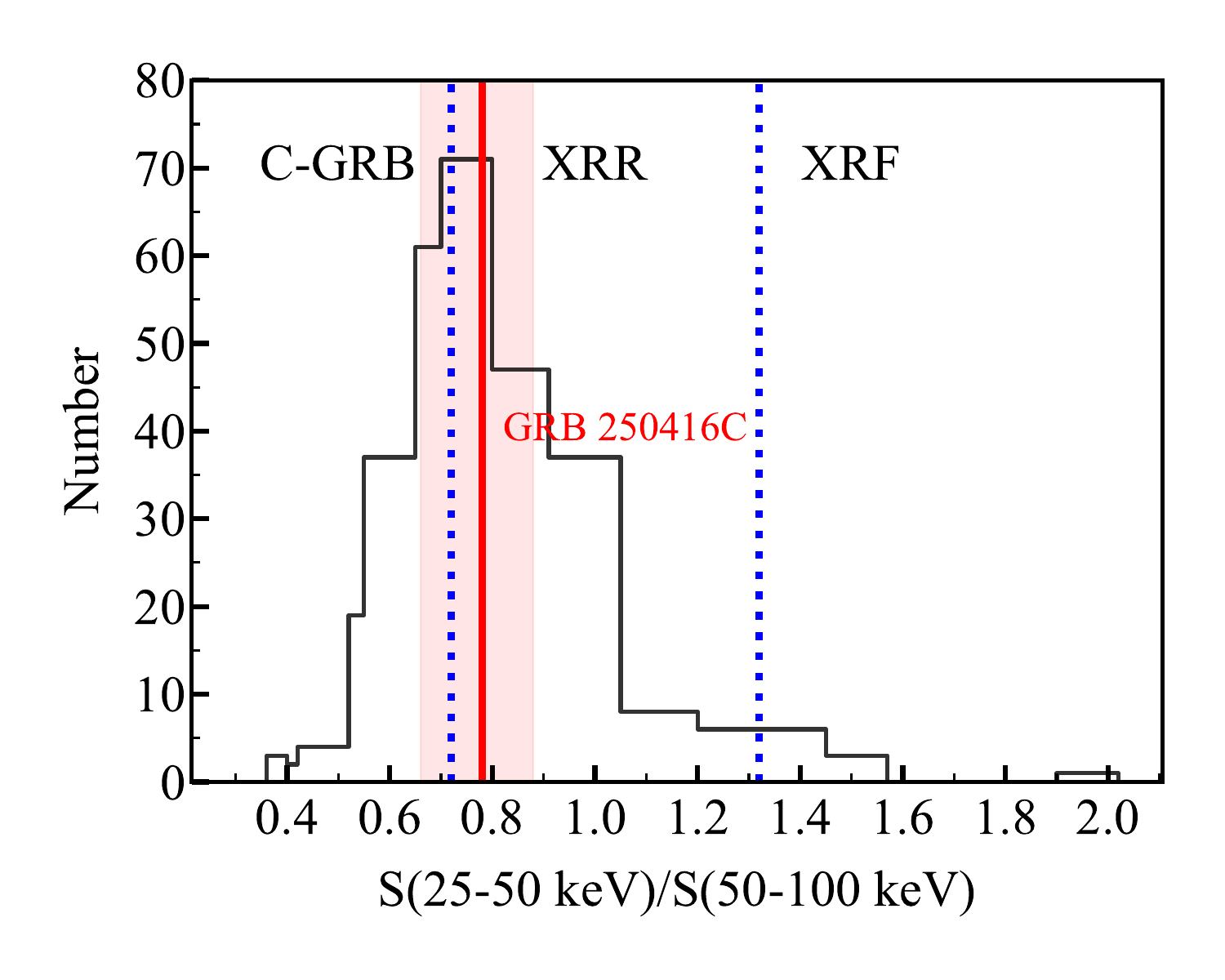}
    \caption{The distribution of the fluence ratio S(25-50 keV)/S(50-100 keV) for the entire sample observed by Swift~\citep{2016ApJ...829....7L}. The shaded region shows the uncertainty of GRB 250416C. The vertical blue dashed lines correspond to the divisions between C-GRBs, XRRs, and XRFs.}
    \label{fig:XRR_XRF_C-GRB}
\end{figure}

For GRB 250416C, the measured fluence ratio is $S(25-50\ \text{keV})/S(50-100\ \text{keV}) = 0.78_{-0.12}^{+0.10}$, While this value is nominally consistent with the XRR subclass, it overlaps with the C-GRB class within the uncertainties, as shown in Figure~\ref{fig:XRR_XRF_C-GRB}. This figure presents the fluence ratio distribution of a comprehensive GRB sample that was compiled by \cite{2016ApJ...829....7L} from primarily Swift detected events. A larger Swift sample covering the ambiguous boundary sources between C-GRBs and XRRs is also reported in ~\cite{2018ApJ...866...97B}. This XRR classification underscores GRB 250416C's distinctive intermediate spectral characteristics: compared to C-GRBs, it exhibits a more prominent X-ray component in its prompt emission; while relative to XRFs, it possesses a stronger $\gamma$-ray emission component. Such a spectral trait further enriches the observed diversity of spectral properties among long GRBs.
 
%%%%%%%%%%%%%%%%%%%%%
%%%%%%%%%%%%%%%%%%%%%

\section{Afterglow emissions}
\label{sec:afterglow_emissions}

Afterglow emission, generated by the interaction between relativistic ejecta from GRBs and the circumburst medium, is critical for probing the burst's energetic scale, jet geometry, and progenitor environment. 

For EP250416a/ GRB 250416C, we analyze multi-wavelength afterglow data from X-ray (EP/FXT, Swift/XRT) and optical facilities, spanning $\rm T_0+142$ s to $\sim \rm T_0+23$ days. The X-ray afterglow of GRB 250416C captures a complete evolutionary sequence, with distinct decay phases and spectral variations that reflect changes in the shock dynamics and energy injection processes. Optical observations, meanwhile, constrain the burst's optical faintness and validate its classification as an optically dark GRB.

\subsection{Temporal Evolution}\label{sec:X-ray follow-up}

The earliest FXT data ($\rm T_0+142$ s to $\sim 1 $hour) captured the transition from the tail of the prompt emission to the onset of the early afterglow, providing key insight into the prompt-afterglow connection. Subsequent observations (1 -23 days) monitored the faint, late-time decay of the source, enabling characterization of the afterglow's long-term temporal evolution. 

To quantify the X-ray flux decay, spectral evolution, and key parameters across all observational epochs, we performed time-resolved spectral fitting for data from WXT, FXT, and Swift/XRT with an absorbed power-law model ($N(E) \propto E^{-\Gamma}$), where $\Gamma$ denotes the photon index. All fits adopt the \texttt{tbabs$\times$powerlaw} model in \texttt{Xspec}. The averaged absorbed flux is derived in the 0.5-4.0 keV band, and uncertainties for all parameters correspond to $1 \sigma$ confidence levels. The fitting results are summarized in Table~\ref{tab:time_resoul}.

%%%%%%%%%%%%%%%%%%%%%
\begin{table*}[]
    \centering
    \renewcommand{\arraystretch}{1.3}
    \caption{Time-resolved spectral fitting results. The averaged absorbed flux is derived in 0.5-4.0 keV band using the spectra model $tbabs\times powerlaw$ for the spectra of WXT, FXT and XRT. All uncertainties quoted here correspond to a $1 \sigma$.}
    \begin{tabular}{*5{c}}
    \hline
    \hline
    Instruments& Time Interval& Photon Index& Flux& Cstat/dof \\
    &[s]& &[$\rm erg~cm^{-2}~s^{-1}$]&\\
    \hline
    \hline
     WXT&(0,30) & $0.35_{-0.15}^{+1.07}$& $(7.7\pm 2.8)\times 10^{-9}$&16.44/23\\
     \hline
     \hline
     &(0.126-1.066)$\times 10^{3}$& $1.75_{-0.31}^{+0.55}$& $(1.5\pm 0.3)\times 10^{-11}$&37.55/35\\
     &(3.666-6.866)$\times 10^{3}$& $2.05_{-0.16}^{+0.16}$& $(8.0\pm 0.5)\times 10^{-12}$&26.89/39\\
     &(4.407-5.007)$\times 10^{4}$& $2.23_{-0.31}^{+0.36}$& $(1.7\pm 0.2)\times 10^{-12}$&37.94/40\\
     FXT&(1.765-1.825)$\times 10^{5}$& $2.80_{-0.62}^{+0.50}$& $(7.5\pm 2.2)\times 10^{-13}$&15.66/27\\
     &(4.414-4.444)$\times 10^{5}$& $2.05_{-0.16}^{+0.16}$& $(1.5\pm 0.7)\times 10^{-13}$&7.86/4\\
     &(6.717-6.808)$\times 10^{5}$& $0.91_{-0.42}^{+0.42}$& $(1.0\pm 1.2)\times 10^{-13}$&7.7/8\\
     &(1.178-1.187)$\times 10^{6}$& $1.28_{-0.76}^{+0.74}$& $(8.0\pm 0.3)\times 10^{-14}$&2.14/3\\
     &(2.487-2.587)$\times 10^{6}$& $0.71_{-0.09}^{+0.09}$& $(1.0\pm 0.8)\times 10^{-14}$&3.98/2\\
     \hline
     \hline
     XRT&(1.449-2.028)$\times 10^{4}$& $1.55_{-0.80}^{+1.18}$& $(3.6\pm 0.5)\times 10^{-12}$&26.19/31\\
     &(7.448-7.748)$\times 10^{4}$& $1.44_{-0.25}^{+0.25}$& $(7.0\pm 1.2)\times 10^{-13}$&19.69/19\\
     \hline
     \hline
    \end{tabular}
    \label{tab:time_resoul}
\end{table*}
  
\begin{figure}[ht!]
\begin{center}
\includegraphics[width=9cm]{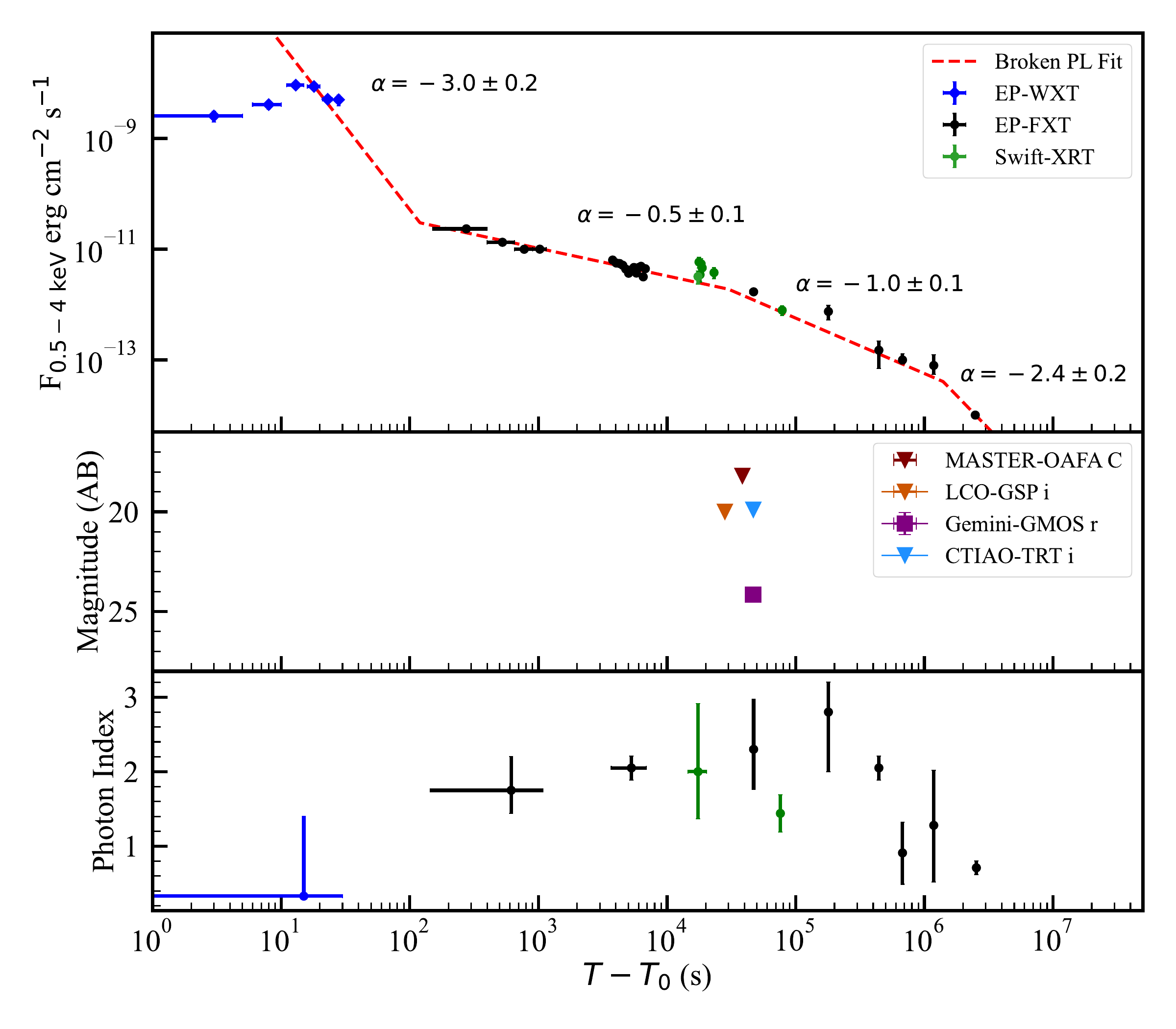} 
\caption{Top panel: X-ray light curve of GRB 250416C. The red dashed line represents the best fit to the data using a broken power-law model. A jet break at t $\sim 1.5\times 10^{6}$ s steepens the decay from $\alpha=-1.0\pm0.1$ to $\alpha=-2.4\pm0.2$. Middle panel: Optical light curve of GRB 250416C. The Gemini-GMOS detection is host-galaxy subtracted. Inverted triangles indicate upper limits. Bottom panel: Evolution of the X-ray spectral photon index $\Gamma$ for GRB 250416C, derived from power-law fits to X-ray spectra.}
\label{fig:X-ray+Optical_lc+spec_evo}
\end{center}
\end{figure}

The X-ray afterglow of GRB 250416C exhibited two distinct decay phases: a shallow-decay phase lasting until t$\sim 2\times 10^{4}$ s with a decay index of $\alpha =-0.5\pm0.1 $, followed by a canonical power-law decay phase from t$\sim 2\times 10^{4}$ s to t$\sim 1.2\times 10^{6}$ s with a decay index of $\alpha =-1.0\pm0.1$. A jet break was identified at t$\sim 1.5\times 10^{6}$ s, where the decay index steepened to $\alpha =-2.4\pm0.2$, characteristic of jet opening angle effects, while no evidence of early jet break signatures were observed in the initial decay phases (Figure ~\ref{fig:X-ray+Optical_lc+spec_evo}, top panel).
        
%%%%%%%%%%%%%%%%%%%%%%
%%%%%%%%%%%%%%%%%%%%%

Spectral evolution is directly quantified by the photon index ($\Gamma$) in Table ~\ref{tab:time_resoul}, revealing a three-stage pattern (Figure~\ref{fig:X-ray+Optical_lc+spec_evo}, bottom panel): an initial softening during the transition from the prompt emission to the shallow decay phase, followed by a period during which the spectrum remains soft with no significant evolution, and finally a hardening phase coinciding with the normal decay and jet break.

%%%%%%%%%%%%%%%%%%%%%%%%
\subsection{Spectral Energy Distribution}\label{sec:SED}

\begin{figure}[ht!]
    \centering
    \includegraphics[width=9cm]{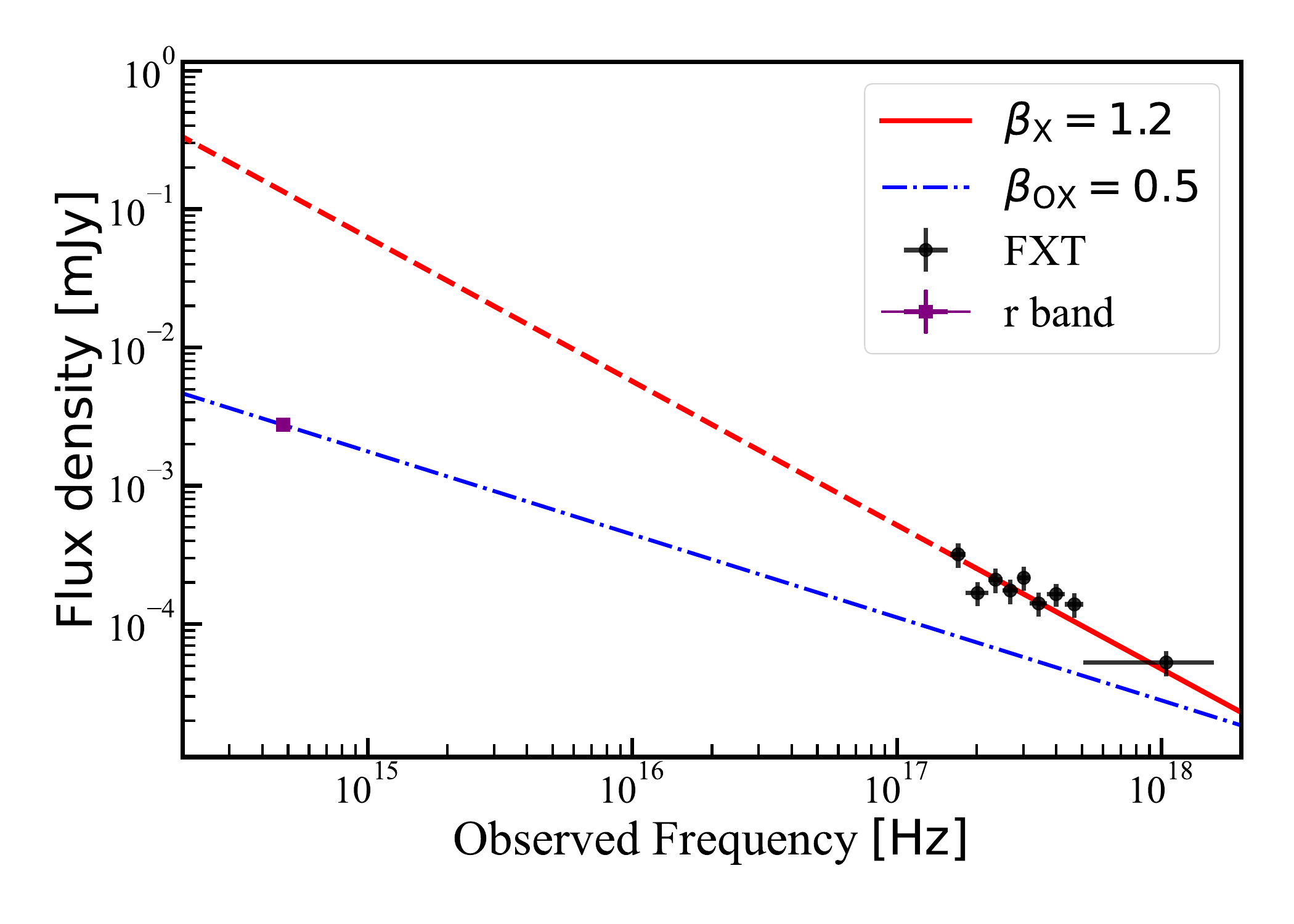}
    \caption{The afterglow spectral energy distribution of GRB 250416C at 47 ks, constructed from the contemporaneous $r$-band and X-ray observations. The solid line shows the best-fitting power-law model obtained from fitting the X-ray spectrum alone, with a spectral index of $\beta_{\rm X}=1.2\pm0.2$. The dashed red line indicates the extrapolation of the X-ray derived power-law model to the optical band, illustrating the deviation of the observed $r$-band flux from the simple power-law expectation. A blue dot-dashed line with a slope of $\beta_{\rm OX}=0.5$ \citep{2004ApJ...617L..21J}, the standard criterion for optically dark GRBs, is also plotted through the $r$-band detection; the location of the X-ray data above this reference suggests an optically dark categorization of this GRB.}
    \label{fig:X-ray_SED}
\end{figure}

A spectral energy distribution (SED) analysis enables a deeper understanding of the afterglow. Figure ~\ref{fig:X-ray_SED} presents the broadband SED at 47 ks, constructed using the contemporaneous $r$-band and X-ray data. We first fit the X-ray spectral data alone with a power-law model of \texttt{Tbabs$\times$zTbabs$\times$PowerLaw} model in \texttt{Xspec}, where \texttt{Tbabs} and \texttt{zTbabs} are the intrinsic equivalent hydrogen column density in the Milky Way Galaxy and the host galaxy, respectively. The best-fitting X-ray powerlaw is shown as the solid line in the figure. The redshift is fixed to 0.963, which yields a host galaxy hydrogen column density of $N\rm_{H,host}=(8.4\pm0.1)\times 10^{21}~ cm^{-2}$ and a spectral index of $\beta_{\rm X}=1.2\pm0.2$. Here the spectral index is defined as $\rm F(E)\propto E^{-\beta_{\rm X}}$ with $\beta_{\rm X} = \Gamma -1$, where $\rm F(E)$ represents the flux density [$\rm erg s^{-1} cm^{-2} keV^{-1}$]. 

Extrapolating this X-ray-derived power-law to the $r$-band (dashed line) reveals that the observed $r$-band flux lies significantly below the model prediction. This discrepancy indicates that the broadband afterglow spectrum at this epoch cannot be described by a single power law, and may be attributed to additional effects such as spectral breaks and/or extinction in the host galaxy. A detailed modeling of the afterglow emission is presented in Section~\ref{sec:modeling}. In accordance with the classification criterion for ``optically dark'' GRBs, events where the optical afterglow is faint or undetectable relative to their X-ray emission~\citep{2004ApJ...617L..21J}, we plot a blue dot-dashed line representing the slope of $\beta_{\rm OX}=0.5$ through the $r$-band detection in Figure~\ref{fig:X-ray_SED}. Its position below the X-ray data confirms the burst to be optically dark. We will discuss this property further in Section~\ref{sec:discussion}.

%%%%%%%%%%%%%%%%%%%%%%%%
\subsection{Comparison with Long GRBs}

\begin{figure}
\includegraphics[width=8.8cm]{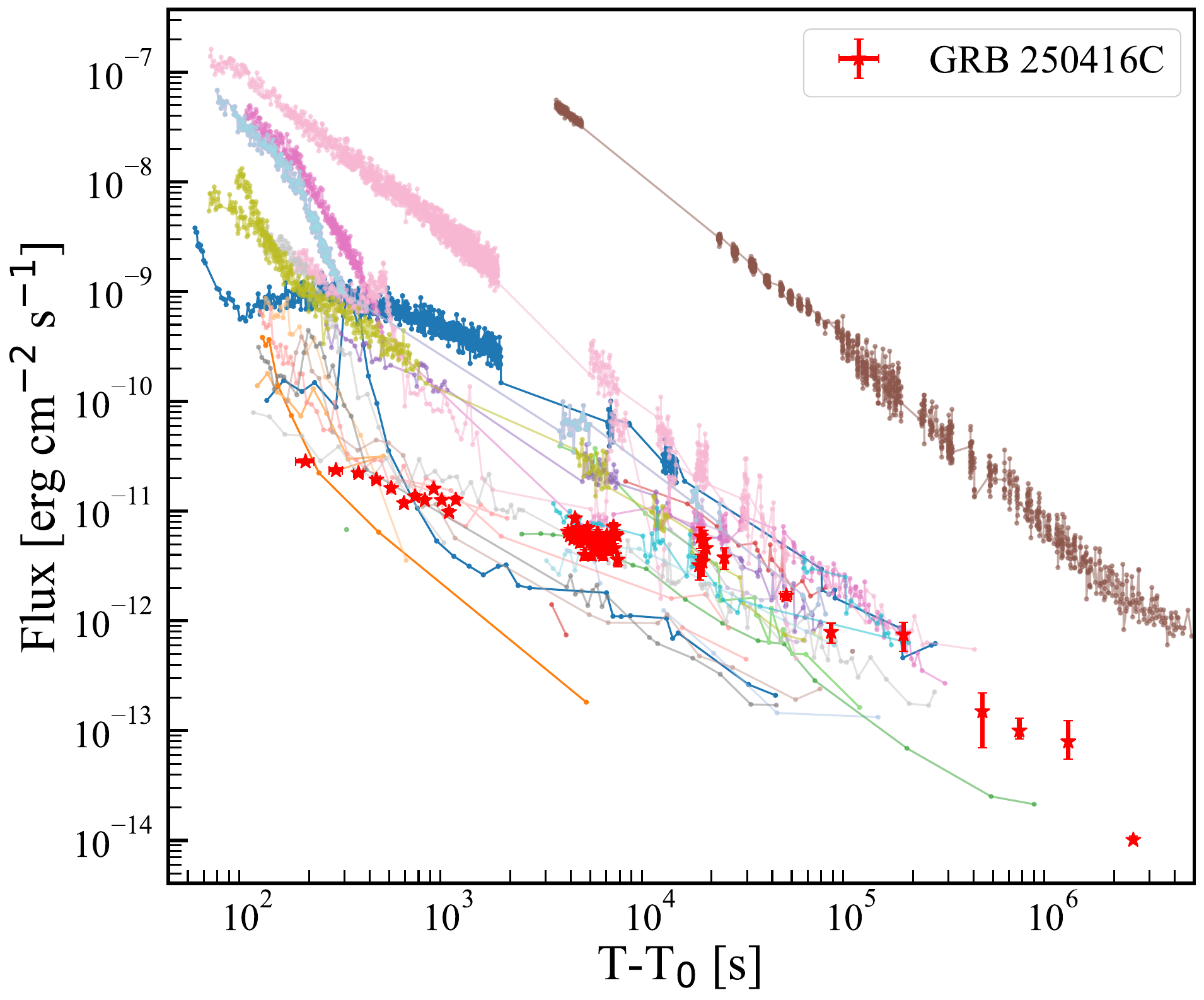}
\includegraphics[width=9cm]{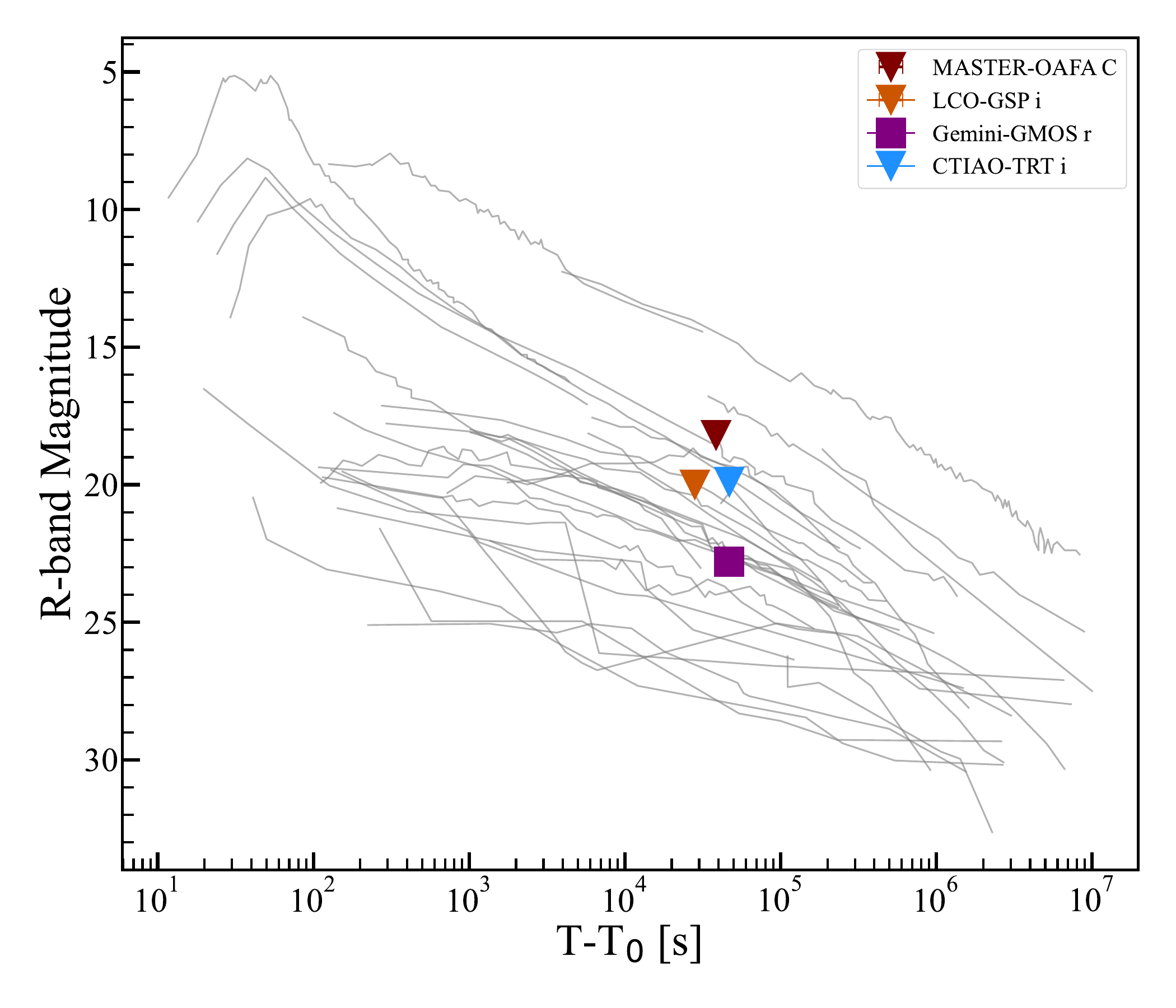}
    \caption{Top: Temporal evolution of 0.3-10.0 keV X-ray flux (observer frame) for Swift-XRT detected long $\gamma$-ray bursts. Public data retrieved from the Swift Data Archive (\url{https://www.swift.ac.uk/xrt_curves/}). GRB 250416C is highlighted in red. Bottom: comparative analysis of the optical luminosity light curve of GRB 250416C with the light curves of optical afterglows from long GRBs. These samples are cited from~\cite{2006ApJ...641..993K,2010ApJ...720.1513K}. In the figure, inverted triangles denote upper limits. }
    \label{fig:Compare_GRBs_sample}
\end{figure}

We compare GRB 250416C's X-ray flux and optical magnitudes to a comprehensive sample of well-studied long GRBs (Figure~\ref{fig:Compare_GRBs_sample}). X-ray data for the comparison sample were retrieved from the Swift Data Archive\footnote{\url{https://www.swift.ac.uk/xrt_curves/}}, covering the 0.3-10 keV band, while optical afterglow luminosities were compiled from previous works by~\cite{2006ApJ...641..993K,2010ApJ...720.1513K}.

%%%%%%%%%%%%%%%%%%%%%%%%%
%%%%%%%%%%%%%%%%%%%%%%%%%%

As highlighted in red in the left panel of Figure~\ref{fig:Compare_GRBs_sample}, the X-ray flux of GRB 250416C falls within the typical range observed for long GRBs. In the optical band, we only detected an $r$-band counterpart with Gemini GMOS, whereas other telescopes placed upper limits on the optical emission.

%%%%%%%%%%%%%%%%%%%%%%%%%%%%%%%%%

%%%%%%%%%%%%%%%%%%%%%%%%%%%%%%%%%%

%%%%%%%%%%%%%%%%%%%%%%%%%%%%%
\section{Afterglow modeling} \label{sec:modeling}

Usually, typical afterglows in GRBs can be well understood within the theory of a relativistic blast wave, i.e., the forward shock (FS) emission produced when a collimated, relativistic ejecta or outflow plunges through the circum-stellar medium and gets decelerated continuously. We first derive rough constraints on the model parameters from several observed afterglow properties. Subsequently, we perform numerical fit to the X-ray and optical afterglow light curves using the model.

 % The jet break time $t_{\rm j}$ provides an estimate of the jet half-opening angle $\theta_{\rm j}$ within the framework of the standard jet afterglow model. In addition, the temporal decay index $\alpha$ during the X-ray normal decay phase and the spectral index $\beta$ can be related to the electron energy distribution index $p$ through the standard afterglow closure relations. These estimates are then adopted to define physically motivated parameter ranges for the subsequent fitting procedure. We use the predictions of this model to numerically fit the X-ray and optical light curves.   
%%%%%%%%%%%%%%%%%%%%%%%%
%\subsection{Decelerating top-hat jet forward-shock model}%

%(Contributor: Duo-le)
\subsection{General Consideration}\label{General Consideration}
In the observer's frame, the jet break time $t_{\rm j}$ is the epoch when the relativistic beaming angle ($1/\Gamma$) becomes comparable to the jet half-opening angle $\theta_{\rm j}$, so the observer begins to see the jet edge. After this time, geometric effects cause the GRB afterglow light curve to decay more steeply. From Figure \ref{fig:X-ray+Optical_lc+spec_evo}, we can directly see that for GRB 250416C the jet break time is $t_{\rm j} \approx 1.5 \times 10^6 \ \rm s$.

The jet half-opening angle can be estimated as (e.g., \citealt{2018pgrb.book.....Z})
\begin{equation}
\theta_{\rm j} \simeq  0.2\ t_{\rm j,6}^{3/8}E_{\rm K,iso,53}^{-1/8} n_{\rm -1}^{1/8} [(1+z)/2]^{-3/8}  \ \rm rad,
\label{Eq:jet break time}
\end{equation}
where the notation $X_{\rm n} \equiv X/10^n$ is used; $E_{\rm K,iso}$ is the isotropic-equivalent total kinetic energy of the jet, and $n$ is the ambient medium number density. The radiative efficiency of a GRB is defined as \citep{2004ApJ...613..477L} $\eta_{\rm \gamma} = E_{\rm \gamma,iso}/(E_{\rm \gamma,iso} + E_{\rm K,iso})$, which is found to vary from less than $1\%$ to over $90\%$ \citep{2007ApJ...655..989Z}. For GRB 250416C,  adopting a typical value of  $\eta_\gamma = 0.2$ and using the other appropriate parameter values, we obtain the half-opening angle of $\theta_{\rm j} \sim 15^{\circ}$. A more accurate estimate is obtained in \S \ref{sec:num_mod}.

Figure \ref{fig:X-ray+Optical_lc+spec_evo} shows that the slope of the X-ray normal decay phase is $\alpha=-1.0\pm 0.1$, and that in this phase (Figure \ref{fig:X-ray_SED}) the spectral index is $\beta_{\rm X} = 1.2\pm0.2$. Assuming a uniform density of the interstellar medium (ISM), the closure relation \citep{2004IJMPA..19.2385Z} suggests that the observed frequency is likely above the cooling frequency $\nu_{\rm c}$ in the slow-cooling regime, or above the minimum injection frequency $\nu_{\rm min}$ in the fast-cooling regime. Meanwhile, the relations between $\alpha$, $\beta_{\rm X}$, and the electron energy distribution index $p$ provide independent constraints on $p$, yielding $p = 4(\alpha+2)/3 = 2.0\pm0.1$ from $\alpha$ and $p = 2\beta_{\rm X} = 2.4\pm0.4$ from $\beta_{\rm X}$.

Finally, we note in Figure \ref{fig:X-ray+Optical_lc+spec_evo} that the X-ray temporal decay slope during $ t= 10^2 \sim  10^4 \ \rm  s$ is $\alpha = -0.5\pm0.1$, which deviates significantly from the slope predicted by the standard afterglow model. Therefore, we consider the possibility that energy might have been injected into the jet due to the activity of the central engine during this period.

\subsection{Numerical Model Fit} \label{sec:num_mod}
Building upon these estimates, we then perform detailed numerical fit to the X-ray and optical afterglow data to refine the parameter constraints. 

To interpret the observed X-ray and optical data throughout the afterglow phase, we employ a relativistic external forward shock (FS) model incorporating energy injection, utilizing the numerical code \texttt{PyFRS}\footnote{\url{https://github.com/leiwh/PyFRS/}} to calculate the multi-band light curve ~\citep{2014ApJ...788...32W,2016ApJ...816...20L,2023ApJ...948...30Z,2024ApJ...963...66Z}. The main physical parameters are: $E_{\rm K,iso}$, $\theta_{\rm j}$, $n$, the jet initial Lorentz factor of $\Gamma_{0}$, the observer viewing angle $\theta_{\rm obs}$, the electron spectral index $p$, the fractions of shock energy that go to electrons $\epsilon_{\rm e}$ and magnetic field $\epsilon_{\rm B}$. 

For the energy injection, we assume that the injection rate follows a power-law form $L(t) = L_{\rm 0} (t/t_{\rm 0})^{-q}$ between the start time $t_{\rm 0}$ and the end time $t_{\rm e}$. The source of energy injection is generally attributed to the activity of the central engine, which may result from the spin-down of a newly born fast-rotating magnetar \citep[e.g.,][]{2001ApJ...552L..35Z} or the fallback accretion onto a stellar-mass black hole (BH) \citep{1977MNRAS.179..433B,2013ApJ...765..125L}.

%The spin-down luminosity of a magnetar depends on its initial parameters. The characteristic spin-down luminosity can be calculated by \citep{2001ApJ...552L..35Z}:
%\begin{equation}
%    L_{\rm 0} = 1.0\times 10^{49}\ (B^2_{\rm p,15} P^{-4}_{\rm 0,-3}R_{\rm 6}^6) \ \rm erg/s 
%\end{equation}
%where $B_{\rm p}$, $P_{\rm 0}$, and $R$ are the magnetic field, spin period, and radius of the magnetar.

%If the central engine is a Kerr BH, its rotational energy can be extracted via the Blandford-Znajek (BZ) mechanism \citep{1977MNRAS.179..433B,2015ApJS..218...12L,2013ApJ...765..125L}.  For a Kerr BH with mass $M_{\rm BH}$ and angle momentum $J_{\rm BH}$, the BZ power can be given by \citep{2000PhR...325...83L,2000PhRvD..61h4016L,2005ApJ...630L...5M,2011ApJ...740L..27L,2013ApJ...765..125L,2017ApJ...849..119C}:
%\begin{equation}
%    L_{\rm BZ} = 1.7\times 10^{50}\ a_{\rm BH}^2m_{\rm BH}^2B_{\rm BH,15}^2F(a_{\rm BH})  \ \rm erg/s
%\end{equation}
%where $a_{\rm BH} = J_{\rm BH}c/(GM^2_{\rm BH})$, $B_{\rm BH}$, and $m_{\rm BH} = M_{\rm BH}/M_{\rm \odot}$ are the spin parameter of the BH, the magnetic-field strength that threads the BH horizon, and the ratio of black hole mass to solar mass, which $M_{\rm \odot}$ denotes solar mass, respectively.
%$F(a_{\rm BH}) = [(1+q_{\rm BH}^2)/q^2_{\rm BH}][(q_{\rm BH}+1/q_{\rm BH})\arctan q_{\rm BH} - 1]$, and $q_{\rm BH} = q_{\rm BH}/(1+\sqrt{1-a_{\rm BH}^2})$.

\begin{table}
    %\centering
    \caption{Parameters of afterglow modeling. Note that $E_\mathrm{K,iso}$ is the initial total isotropic equivalent-kinetic energy of the jet, excluding the late-time injected energy.}
    \renewcommand{\arraystretch}{1.2}
    \begin{tabular}{cccc}
    \hline 

    \hline
    Parameters& Prior Type& Prior& Results \\
    \hline
    $E_\mathrm{K,iso}$ [erg]& log-uniform& [$10^{52}$, $10^{56}$]& $(3.9_{-1.3}^{+2.0}) \times 10^{52}$ \\
    $\Gamma_{\rm 0}$& log-uniform& [10,1000]&$(3.0_{-1.2}^{+1.3})\times 10^2$\\
    $p$& uniform& [2,3]&$2.3_{-0.1}^{+0.2}$\\
    $n$ [cm$^{-3}$]& log-uniform& [$10^{-4}$,100]&$0.3_{-0.1}^{+0.2}$ \\
    $\theta_{\rm j}$ [degree]& uniform&[0.1,25]&$10.6_{-1.8}^{+1.9}$ \\
    $\theta_\mathrm{obs}$ [degree]& fixed&0 &0 \\
    $L_{\rm 0} [\rm{erg/s}$]&log-uniform &$[10^{45},10^{54}]$ &$(3.3_{-1.3}^{+1.9})\times 10^{49}$\\
    $t_{\rm 0}$ [\rm{s}]&log-uniform &$[1,10^{3}]$ &$(4.7_{-1.1}^{+1.3})\times 10^{2} $\\
    $t_{\rm e}$ [\rm{s}]&log-uniform &$[10^{3},10^{6}]$ &$(2.9_{-1.2}^{+2.8})\times 10^{4}$\\
    $q$& log-uniform&[$10^{-2}$,2] &$0.4_{-0.1}^{+0.1}$\\
    $\epsilon_{\rm e}$& log-uniform& [$10^{-4}$, 0.5] &$(2.6_{-0.9}^{+1.2})\times 10^{-2}$\\
    $\epsilon_{\rm B}$& log-uniform & [$10^{-4}$, 0.5] &$(8.7_{-2.9}^{+5.2}) \times 10^{-4}$\\
    \hline 

    \hline
    \end{tabular}
    \vspace{0.3cm}
    
    \label{tab:X-ray_FS_SED}
\end{table}

\begin{figure}[ht!]
    \centering
    \includegraphics[width=9cm]{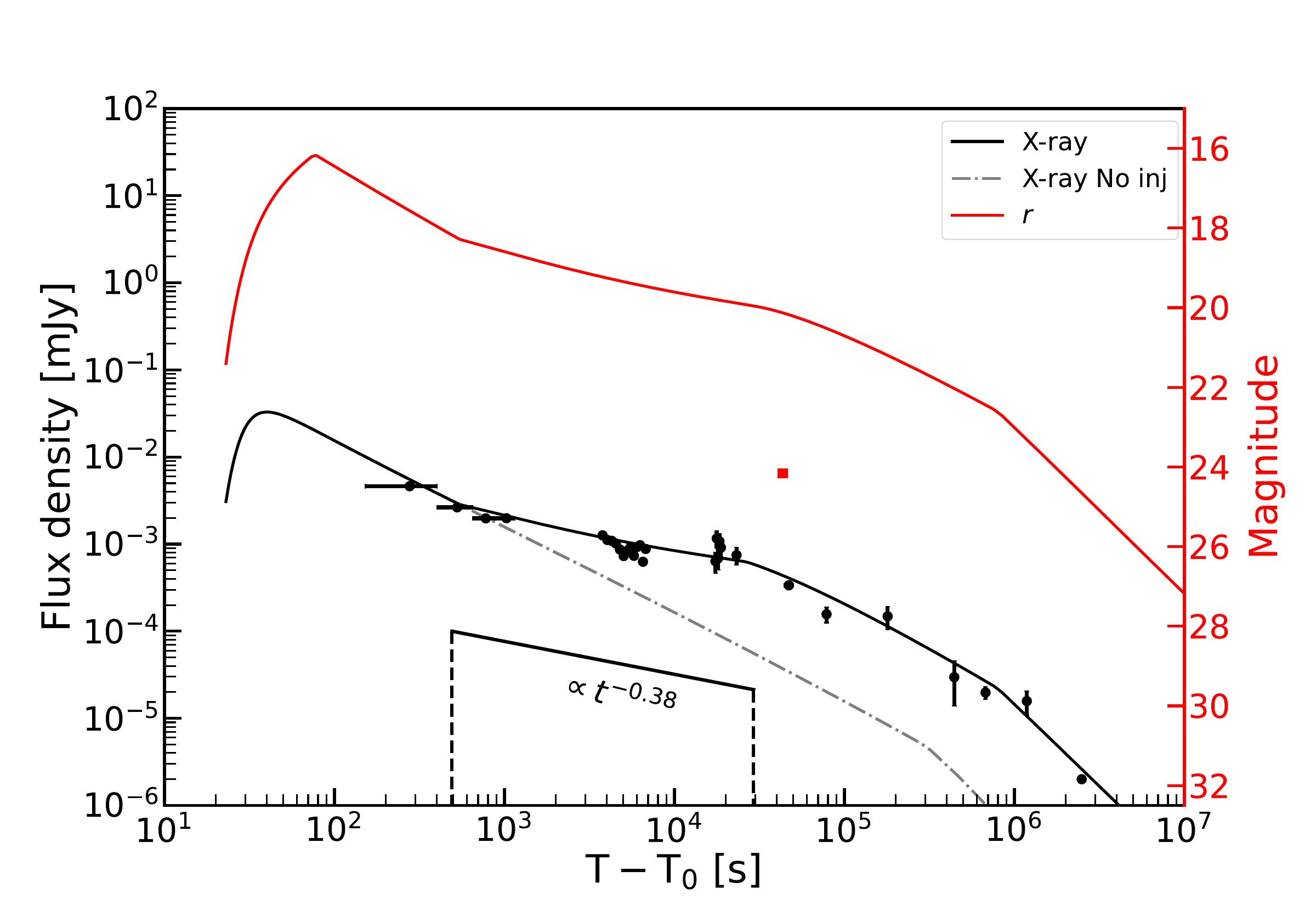}
    \caption{The best-fit light curve for GRB 250416C X-ray and optical data. 
Black circles and red squares represent the data in the X-ray band and the optical \textit{r}-band, respectively. The black solid line represents the fitted X-ray light curve obtained using the \texttt{PyFRS} code. The red solid line shows the $r$-band light curve calculated using the best-fit parameters derived from the X-ray data. FS without energy injection (No inj) are represented by the black dash-dotted line. The black dashed line represents the temporal evolution of the energy injection rate, and the starting and ending points of the curve correspond to the onset ($t_{\rm 0}$)  and cessation ($t_{\rm e}$) of the energy injection, respectively.}
    \label{fig:multi-band light curve energy injection}
\end{figure}

We utilize the Python package \texttt{emcee} to perform Markov Chain Monte Carlo (MCMC) sampling for deriving the posterior distributions of afterglow parameters. Log-uniform priors are adopted for $E_{\rm K,iso}$, $\Gamma_{\rm 0}$, $n$, $\epsilon_{\rm e}$, $\epsilon_{\rm B}$, while uniform priors are assigned to $p$ and $\theta_{\rm j}$. The parameter space is explored through 60,000 MCMC steps, with the initial $30\%$ of the steps discarded as burn-in. When the X-ray and optical data are fitted simultaneously, the resulting fits for both bands are unsatisfactory, likely due to the optical dimness of this burst. Since only a single data point is available in the optical band, we therefore choose to fit the X-ray data alone. The prior ranges and estimates of the best-fit parameters are tabulated in Table~\ref{tab:X-ray_FS_SED}, with the light curve shown in Figure~\ref{fig:multi-band light curve energy injection} and the corresponding posterior probability distributions in Figure~\ref{appfig:corner}.

In Figure \ref{fig:multi-band light curve energy injection}, the black circles and red squares represent the data in the X-ray band and the optical \textit{r}-band, respectively. The solid black line represents the fitted X-ray light curve obtained using the \texttt{PyFRS} code. The red purple line shows the $r$-band light curve calculated using the best-fit parameters derived from the X-ray data. As can be seen in the figure, the $r$-band data point is significantly lower than the predicted value. This discrepancy may be caused by strong extinction in the host galaxy, as will be discussed in detail in \S \ref{sec:discussion}. 

The black dashed line in Figure \ref{fig:multi-band light curve energy injection} represents the temporal evolution of the energy injection rate. For comparison, the FS without energy injection (``No inj'') is represented by the black dashed-dotted line. The model light curve is flattened during the energy injection interval, and the jet-break time is correspondingly delayed. 

Although this blastwave-with-injection model provides a reasonable description of the observed features, we note that due to limited data coverage in the optical band and those in missing during the early time ($t\lesssim 10^2~\rm s$), it is merely one plausible solution and not the unique explanation for the multi-band data.

The constraint on the jet initial Lorentz factor $\Gamma_{\rm 0}$ is rather weak, as can be seen in Figure \ref{appfig:corner}, with a tendency to higher values. This is because $\Gamma_{\rm 0}$ is primarily determined by the peak time $t_{\rm peak}$ of the afterglow light curve, corresponding to the deceleration time of the jet \citep{1999ApJ...520..641S,2010ApJ...725.2209L}, as in
\begin{equation}
    \Gamma_{\rm 0} \simeq 340\ t_{\rm p,z,2}^{-3/8}E_{\rm K,iso,53}^{1/8} n_{\rm -1}^{-1/8}
\end{equation}
where, $t_{\rm p,z} = t_{\rm peak}/(1+z)$. For GRB 250416C, $t_{\rm peak}$ was not detected, suggesting that  $t_{\rm peak} \lesssim 200\ \rm s$. Adopting the appropriate values for other parameters, then one finds $\Gamma_{\rm 0} \gtrsim 300$. This agrees with the best-fit $\Gamma_{\rm 0}$ listed in Table  \ref{tab:X-ray_FS_SED}. 

\begin{figure}[ht!]
    \centering
    \includegraphics[width=9cm]{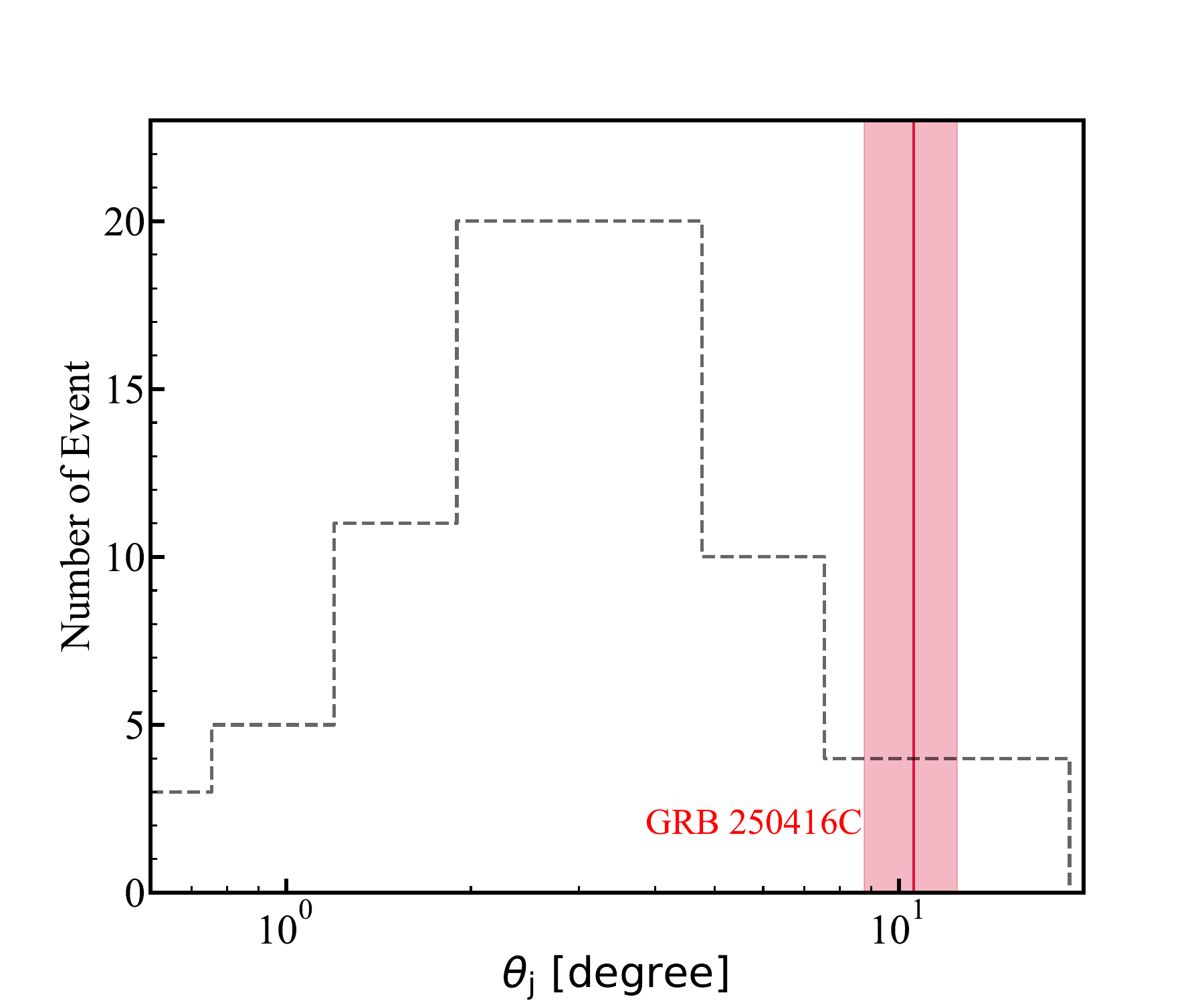}
    \caption{Distribution of the jet opening angle for GRBs from \cite{2012ApJ...745..168L}. The red shaded region corresponds to $\theta _{\rm j}$ of GRB 250416C.}
    \label{fig:Theta_j_Distribution}
\end{figure}

In addition,  we plot $\theta _{\rm j}$ of GRB 250416C against the distribution of jet half-opening angles for a GRB sample from \cite{2012ApJ...745..168L} in Figure ~\ref{fig:Theta_j_Distribution}. It shows that the jet of GRB 250416C is extremely wide, compared to most GRBs.

\begin{figure}[ht!]
\includegraphics[width=9cm]{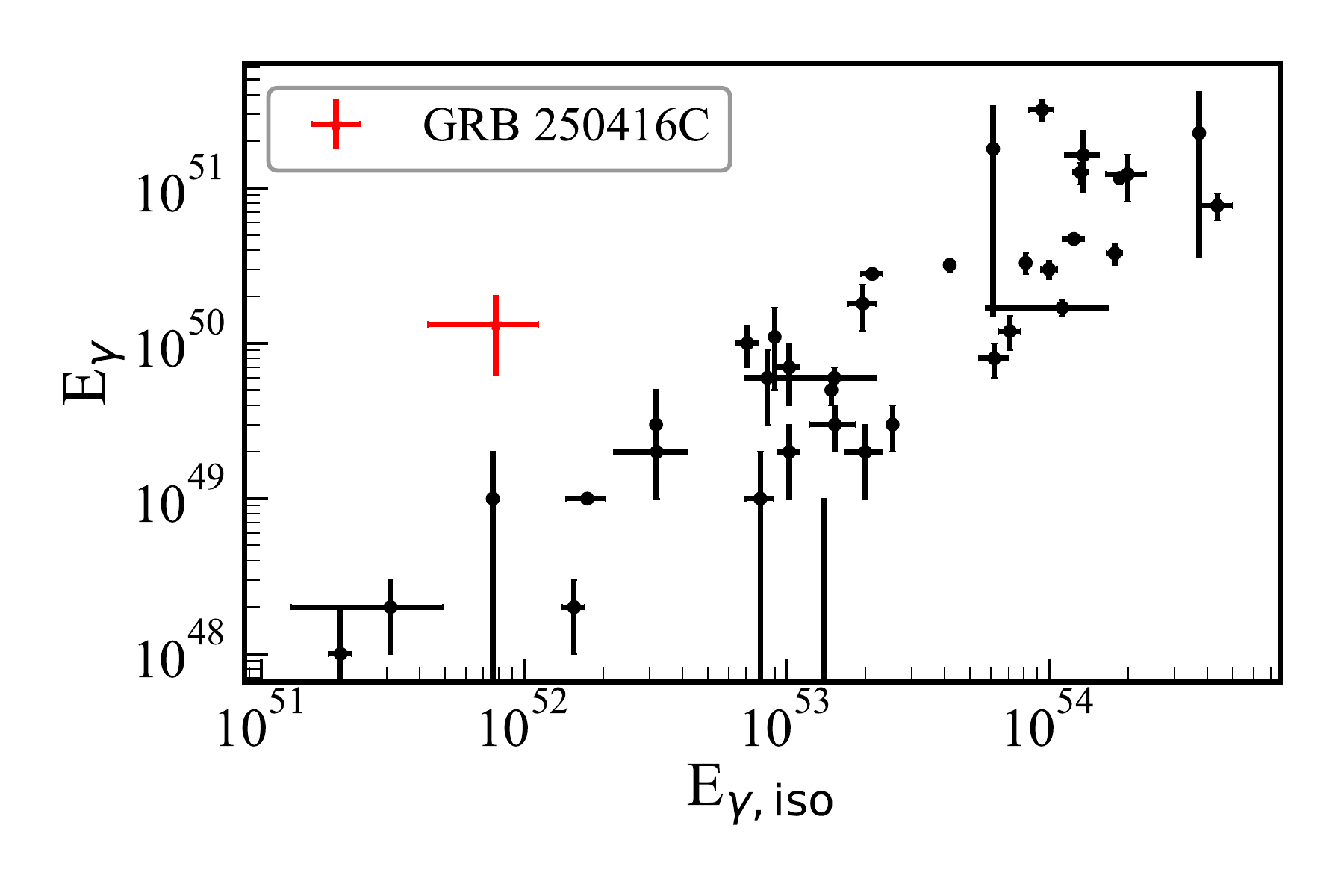}
\includegraphics[width=9cm]{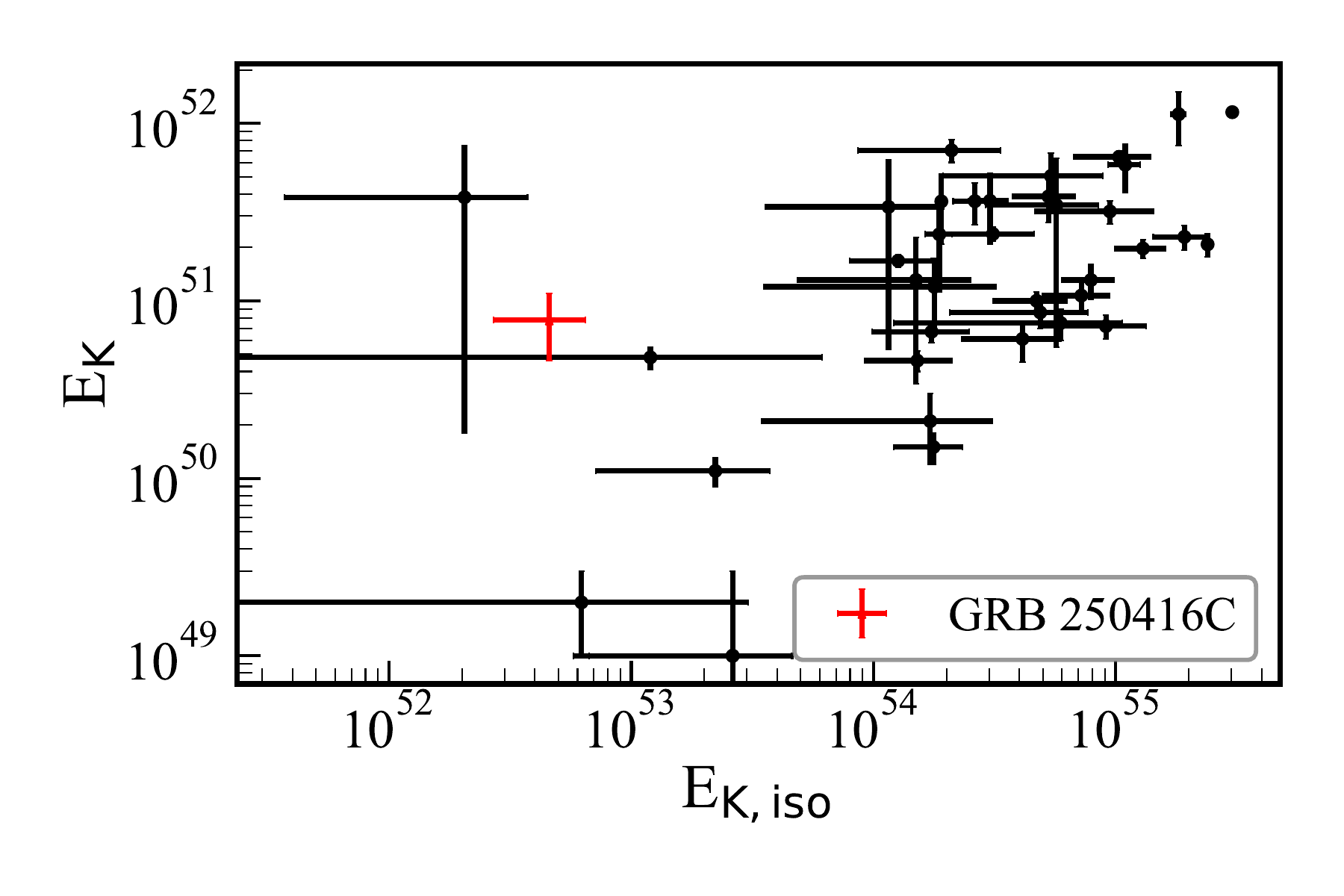}
    \caption{The distribution of GRBs in the E$_{\rm \gamma,\rm iso}$(E$_\mathrm{K,iso}$)-E$_{\gamma}$(E$_\mathrm{K}$) plane. The gray dots denote the GRB sample complied by \cite{2018ApJ...859..160W}, covering both LGRBs and SGRBs to reflect the overall energy output characteristics of GRBs. The red ``+'' symbol hightlights the position of GRB 250416C.}
    \label{fig:E-Eiso}
\end{figure}

To further investigate GRB 250416C's energetics with respect to the broader GRB population, we plot its $E_{\rm \gamma,iso}$ and $E_{\rm K,iso}$, respectively, versus their beaming-corrected energies $E_{\gamma} (E_\mathrm{K})\simeq \theta_\mathrm{j}^2/2 \times E_\mathrm{\gamma,iso} (E_\mathrm{K,iso})$ in Figure \ref{fig:E-Eiso}. We overlay GRB 250416C onto the distributions for a GRB sample compiled by \cite{2018ApJ...859..160W}. It shows that GRB 250416C appears to lie above  the $E_{\rm \gamma}$-$E_{\rm \gamma,iso}$ relation of the sample, while its deviation of $E_\mathrm{K}$-$E_{\rm K,iso}$ relative to that of the sample is less, apparently due to the large uncertainties of $E_\mathrm{K} (E_{\rm K,iso})$ of the sample. The position of GRB 250416C seems to reflect the relatively large $\theta_{\rm j}$ of this burst (cf. Figure \ref{fig:Theta_j_Distribution}). Again, we caution that this result is based on the blastwave-with-injection model, which might not be the unique explanation for this burst.

%%%%%%%%%%%%%%%%%%%%%%%%%%%%%%%

%%%%%%%%%%%%%%%%%%%%%%%%%%%%%%%
\section{Optical darkness}
\label{sec:discussion}

%\subsection{Optical darkness}\label{sec:optical darkness}

GRB 250416C was only detected in the optical band by the Gemini telescope. Two criteria have been proposed to classify GRBs as being optically dark. The first uses simultaneous optical and X-ray fluxes measured approximately half a day after the GRB onset and requires that $\beta \rm_{OX}<0.5$~\citep{2004ApJ...617L..21J}. Here, the optical-to-X-ray spectral index is defined \citep{2025arXiv251215162L} as $\beta \rm_{OX}=-\log (F_{\nu_X} /F_{\nu_o}) / \log (\nu_{\rm X} / \nu_{\rm o})$, where $F_{\rm \nu_X}$, $F_{\rm \nu_{\rm o}}$ are the X-ray / optical flux densities, respectively. We adopt the X-ray flux density at 1 keV from the contemporaneous spectrum and the host-subtracted $r$-band optical flux density for this calculation. The second depends solely on spectral data, with the criterion $\beta \rm_{OX}<\beta \rm_{X}-0.5$ (where $\beta \rm_{X}$ is the X-ray spectral index;~\cite{2009ApJ...699.1087V}). From the simultaneous multi-wavelength data, we derived $\beta_{\rm OX}=0.3\pm0.1$ for GRB 250416C. This value satisfies both of the above criteria, given $\beta \rm_{X}=1.2\pm0.2$ (see \S~\ref{sec:SED}). These securely classify GRB 250416C as an optically dark burst. 

The clear optical darkness of GRB 250416C raises the question of its physical origin. A natural explanation is the significant dust extinction within its host galaxy, which could suppress the observed optical emission. We notice in Figure~\ref{fig:multi-band light curve energy injection} that, while the X-ray afterglow is well described by the forward shock model, the observed optical flux lies significantly below the model prediction. If it were due to host dust extinction, one can infer a required $r$-band extinction of $A(r)=4.16$~mag. The central wavelength of observed GMOS $r$-band is around 7317\text{\AA}, corresponding to a rest-frame wavelength of $\sim 3727\text{\AA}$ at redshift $z=0.963$. Hence the derived extinction characterizes $A_{3727\text{\AA}}$ instead of rest-frame $r$-band extinction. Assuming the standard Milky Way extinction law for the host $A(\lambda)/A(V)=a(x)+b(x)/R_V$ with R$\rm _V=3.1$, and adopting coefficients $a(x)=0.953$ and $b(x)=1.909$ for the 3727\text{\AA} \citep{1989ApJ...345..245C}, one can infer a required $V$-band extinction of $A(V)\approx 2.7\pm 0.3$ mag.

To corroborate this scenario, we utilize the independent constraint from X-ray absorption measurements and employ an empirical relation between X-ray absorbing column density and optical extinction $N_{\rm H}$  (cm$^{-2}$) $= (2.21\pm 0.09)\times 10^{21}~A_{\rm V}$ \citep{1975ApJ...198...95G, 1995A&A...293..889P, 2009MNRAS.400.2050G}. Adopting and the host column density $N_{\rm H,host} =(8.4\pm 0.07)\times 10^{21}$ cm$^{-2}$ obtained in \S \ref{sec:SED}, we derive $A\rm _V^{host}=3.8\pm 0.2$ mag. 

Given the inherent uncertainties associated with the derivations on either side, this value is broadly consistent with the extinction of \textbf{$A(V)\approx 2.7\pm 0.3$}~mag derived from the optical flux deficit, supporting a high degree of dust extinction in the host galaxy.

In Table~\ref{tab:A_V}, we list the inferred host extinctions of 8 dark GRBs with $z<1$, which shows that their $A_{\rm V}$'s cover a wide range. The extinction derived for GRB 250416C is notably higher than the typical range for such populations. 

\begin{table}[]
    \centering
    \caption{Required host galaxy extinction values for several optically dark GRBs.}
    \begin{tabular}{cccc}
    \hline
         GRB name& $z$ &$A\rm_V^{host}(mag)$ &Reference  \\
         \hline
         970828& 0.958 &$>3.8$ & \cite{2001ApJ...562..654D}\\
         000210& 0.85& 0.9-3.2& \cite{2002ApJ...577..680P}\\
         020809&0.41& 0.6-1.5 &\cite{2005ApJ...629...45J}\\
         051022&0.809&$>2.3$&\cite{2007ApJ...669.1098R}\\
         050416A&0.654& 3.8&\cite{2007AJ....133..122H}\\
         140713A & 0.935& $>3.2$ &\cite{2019MNRAS.484.5245H}\\
         240825A& 0.659& 2.9 &\cite{2025arXiv251215162L}\\
         250416C &0.963 & $2.7\pm 0.3$ & This work\\
         \hline
    \end{tabular}
    \label{tab:A_V}
\end{table}

%%%%%%%%%%%%%%%%%%%%%%%%%%%%%%%%%%
\section{Conclusion}
\label{sec:conclusion}

We present a comprehensive multi-wavelength study of EP250416a/GRB 250416C, triggered by EP/WXT on 2025 April 16. By integrating data from EP, Swift, Konus-Wind, SVOM-GRM and ground-based optical telescopes, we systematically characterize its prompt emission, afterglow evolution, and physical origin. Our conclusions are summarized as follows:
\begin{itemize}
    \item 
    GRB 250416C is unambiguously classified as a long-duration GRB, with a $\gamma$-ray band duration T$_{90,\gamma}=17.7$ s in the 18-75 keV and 75-311 keV bands, which falls well within the criterion for long GRBs (T$_{90}\geq 2$ s). It further exhibits a spectral property consistent with XRR subclass, although the fluence ratio S(25-50 keV)/S(50-100 keV) = $0.78_{-0.12}^{+0.10}$ overlaps the C-GRB boundary within uncertainties. It's position on the Amati relation aligns with typical long GRBs, confirming consistency with the empirical spectral-energy correlation of this GRB class.
    
    \item The X-ray afterglow of GRB 250416C exhibits a two-phase decay profile: an initial shallow decay phase ($\alpha=-0.5\pm0.1$) lasting until $t\sim 2\times 10^4$ s, followed by a possible canonical power-law decay phase ($\alpha=-1\pm0.1$) extending to $t\sim 1.2\times 10^6$ s. A jet break is identified at $t\sim 1.6\times 10^6$ s, where the decay index steepens to $\alpha=-2.4\pm 0.2$. This transition is a key signature of relativistic beaming effects, enabling the estimation of the jet's physical opening angle.

    \item A forward shock model with energy injection was employed to fit the afterglow data, verifying that the emission aligns with the relativistic forward shock paradigm. Key derived parameters reveal that the initial isotropic-equivalent kinetic energy of the jet is $E\rm_{K,iso}=(3.9_{-1.3}^{+2.0})\times 10^{52}~erg$, and its half-opening angle is $10.6_{-1.8}^{+1.9}$ degrees, which is significantly wider than the jets in most GRBs. 
    
    \item Notably, the X-ray light curve slope of $\simeq -0.5\pm0.1$ over the $10^2-10^4$ s interval deviates substantially from the standard afterglow model prediction. This discrepancy can be explained by a central-engine-driven energy injection, which may originate from either the magnetar spin-down or the black hole Blandford-Znajek mechanism. The total injected energy is $\simeq 3\times 10^{53}~\rm erg$, significantly larger than $E_\mathrm{K,iso}$, with a power-law decaying energy injection rate characterized by a temporal index $q\simeq0.4$.

    \item This burst is classified as an optically dark GRB: most ground-based telescopes only constrain upper limits on optical flux, with only Gemini South-GMOS detecting a faint $r$-band afterglow of 24.2 mag. We infer that a host galaxy extinction of \textbf{$A_V\approx 2.7\pm 0.3$} mag is required to explain the observed optical deficit. This substantial dust extinction is the primary cause of its optical faintness.
    
\end{itemize}

This study highlights the value of coordinated multi-wavelength observations for unraveling GRB physics. GRB 250416C's is an optically dark long GRB enriches our understanding of GRB diversity, emphasizing the role of host extinction in optical faintness and central engine energy injection in afterglow evolution.

\begin{acknowledgements}
This work is supported by National Natural Science Foundation of China (NSFC-12393814) and by National Key R\&D Program of China No. 2025YFF0511100. D.S., D.F., A.R., A.L., and M.U. was supported by the basic funding program of the Ioffe Institute no. FFUG-2024-0002. P.G.J. and J.Q.V. are supported by the European Union (ERC, Starstruck, 101095973, PI Jonker). Views and opinions expressed are, however, those of the author(s) only and do not necessarily reflect those of the European Union or the European Research Council Executive Agency. Neither the European Union nor the granting authority can be held responsible for them. DBM is funded by the European Union (ERC, HEAVYMETAL, 101071865). The Cosmic Dawn Center (DAWN) is funded by the Danish National Research Foundation under grant DNRF140. FEB acknowledges support from ANID-Chile BASAL CATA FB210003 and FONDECYT Regular 1241005.
\end{acknowledgements}

\bibliography{reference}{}
\bibliographystyle{aa}

\onecolumn
\appendix
\section{}

\begin{table*}[!htbp]
\centering
\begin{threeparttable}
\caption{Optical photometric observations of GRB 250416C/EP250416a.}
\label{apptable:Optical Photometric Observations}
\begin{tabular}{cccc}
\hline
\hline
$T-T_0$ (s) & Telescope/Instrument & Band & AB magnitude \\
\hline
$28800$    & Las Cumbres Observatory 1~m & $i$ & $>20$\textsuperscript{a} \\
37127  & MASTER-OAFA          & Clear ($C$) & $>16.5$\textsuperscript{b} \\
37187  & MASTER-OAFA          & Clear ($C$) & $>17.7$\\
37590   & MASTER-OAFA          & Clear ($C$) & $>17.3$ \textsuperscript{b}\\
38093   & MASTER-OAFA          & Clear ($C$) & $>17.5$\textsuperscript{b} \\
38511   & MASTER-OAFA          & Clear ($C$) & $>17.7$\textsuperscript{b} \\
39442   & MASTER-OAFA          & Clear ($C$) & $>18.2$ \textsuperscript{b}\\
$46656$        & Gemini-GMOS & $r$ & $22.8\pm0.1$\textsuperscript{c}\\
$46800$     &CTIAO-TRT & $g$ & $>21.1$ \\
$46800$        &CTIAO-TRT & $r$ & $>21.0$ \\
$46800$     &CTIAO-TRT & $i$ & $>19.9$ \\
$123264$       & NOT/ALFOSC & $r$ & $23.42\pm0.22$ \\
\hline
\hline
\end{tabular}
\begin{tablenotes}
\footnotesize 
\item[ ]Filter $C$ corresponds to the unfiltered (clear) band of MASTER-OAFA.
${\mathrm{^{a}}}$\cite{2025GCN.40157....1L}. ${\mathrm{^{b}}}$\cite{2025GCN.40156....1L}. ${\mathrm{^{c}}}$\cite{2025GCN.40160....1L}
\end{tablenotes}
\end{threeparttable}
\end{table*}

\begin{figure*}[!ht]
    \centering
    \includegraphics[width=15.3cm]{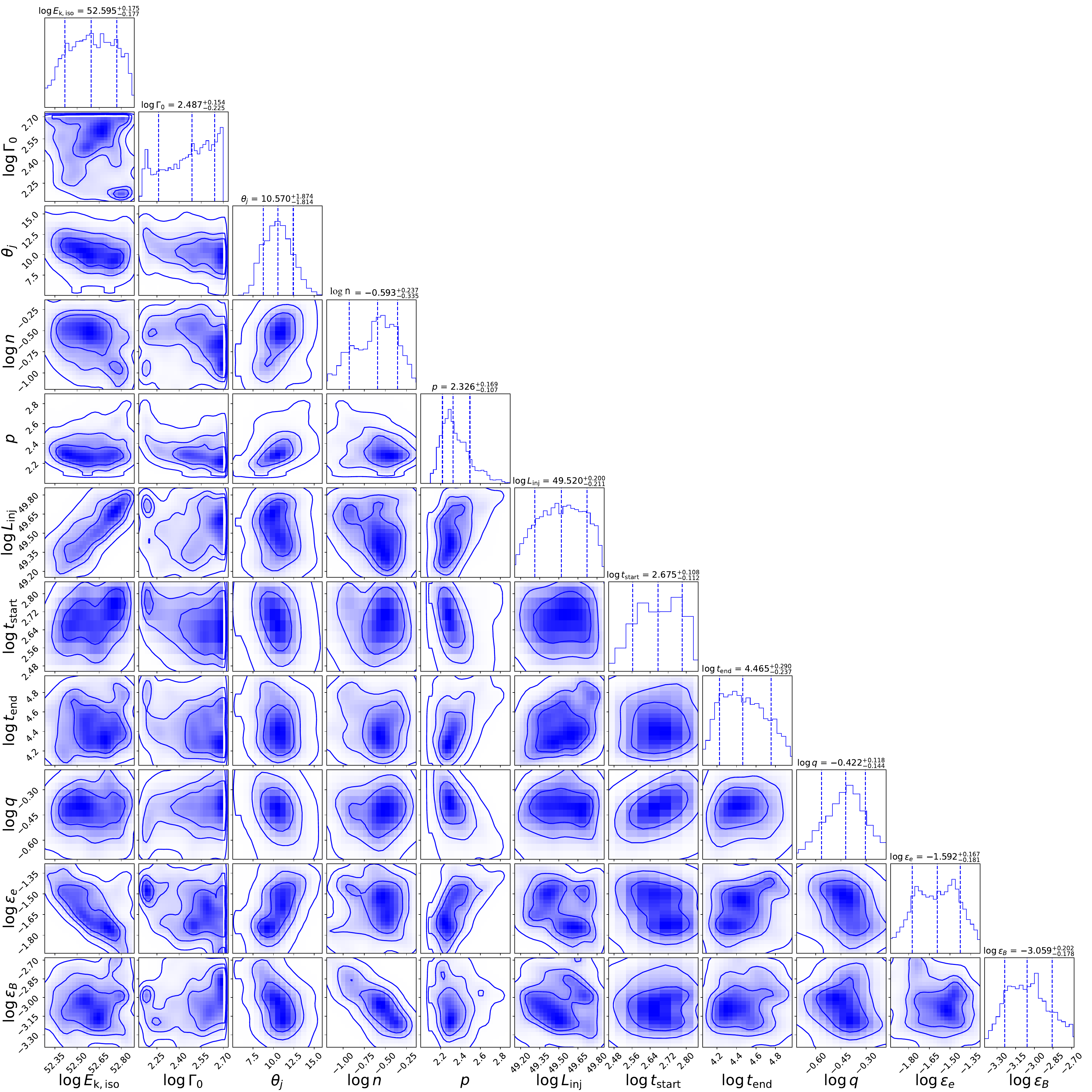}
    \caption{Posterior distributions of afterglow parameters for GRB 250416C.}
    \label{appfig:corner}
\end{figure*}

\end{CJK*}
\end{document}